\documentclass[12pt,draftclsnofoot,onecolumn]{IEEEtran}
\usepackage[T1]{fontenc}
\ifCLASSINFOpdf
\else
\fi
\usepackage{amsmath,amssymb,amsfonts}
\usepackage{diagbox}
\usepackage{bbding}
\usepackage[cmintegrals]{newtxmath}
\usepackage[utf8]{inputenc}
\usepackage{babel}
\usepackage{multirow}
\usepackage{graphicx}
\usepackage{cite}
\usepackage{subfigure}
\usepackage{threeparttable}
\usepackage{booktabs}
\usepackage[table ]{ xcolor}
\definecolor{mygray}{gray}{.7}
\begin{document}
\renewcommand\arraystretch{1}
\bibliographystyle{ieeetr}
\title{\huge Reconfigurable Intelligent Surface-Enhanced OFDM Communications via Delay Adjustable Metasurface}
\author{Jiancheng~An, Chao~Xu,~\emph{Senior~Member,~IEEE}, Chau~Yuen,~\emph{Fellow,~IEEE}, Derrick~Wing~Kwan~Ng,~\emph{Fellow,~IEEE}, Lu~Gan, and Lajos~Hanzo,~\emph{Fellow,~IEEE}
	\thanks{L. Hanzo would like to gratefully acknowledge the financial support of the Engineering and Physical Sciences Research Council projects EP/N004558/1, EP/P034284/1, EP/P003990/1 (COALESCE), of the Royal Society’s Global Challenges Research Fund Grant as well as of the European Research Council’s Advanced Fellow Grant QuantCom. J. An and L. Gan are with the School of Information and Communication Engineering, University of Electronic Science and Technology of China (UESTC), Chengdu, Sichuan 611731, China. L. Gan is also with the Yibin Institute of UESTC, Yibin, Sichuan 644000, China. (e-mail: jiancheng$\_$an@163.com; ganlu@uestc.edu.cn). C. Xu and L. Hanzo are with the School of Electronics and Computer Science, University of Southampton, Southampton SO17 1BJ, U.K. (e-mail: cx1g08@soton.ac.uk; lh@ecs.soton.ac.uk). C. Yuen is with the Engineering Product Development (EPD) Pillar, Singapore University of Technology and Design, Singapore 487372 (e-mail: yuenchau@sutd.edu.sg). D. W. K. Ng is with the School of Electrical Engineering and Telecommunications, University of New South Wales, Sydney, NSW 2052, Australia (e-mail: w.k.ng@unsw.edu.au).}
	}
\markboth{DRAFT}{DRAFT}
\maketitle
\vspace{-2cm}
\begin{abstract}
Reconfigurable intelligent surface (RIS) is a promising technology for establishing spectral- and energy-efficient wireless networks. In this paper, we study RIS-enhanced orthogonal frequency division multiplexing (OFDM) communications, which generalize the existing RIS-driven context focusing only on frequency-flat channels. \emph{Firstly}, we introduce the delay adjustable metasurface (DAM) relying on varactor diodes. In contrast to existing reflecting elements, each one in DAM is capable of storing and retrieving the impinging electromagnetic waves upon dynamically controlling its electromagnetically induced transparency (EIT) properties, thus additionally imposing an extra delay onto the reflected incident signals. \emph{Secondly}, we formulate the rate-maximization problem by jointly optimizing the transmit power allocation and the RIS reflection coefficients as well as the RIS delays. Furthermore, to address the coupling among optimization variables, we propose an efficient algorithm to achieve a high-quality solution for the formulated non-convex design problem by alternately optimizing the transmit power allocation and the RIS reflection pattern, including the reflection coefficients and the delays. \emph{Thirdly}, to circumvent the high complexity for optimizing the RIS reflection coefficients, we conceive a low-complexity scheme upon aligning the strongest taps of all reflected channels, while ensuring that the maximum delay spread after introducing extra RIS delays does not exceed the length of the cyclic prefix (CP). \emph{Finally}, simulation results demonstrate that the proposed design significantly improves the OFDM rate performance as well as the RIS's adaptability to wideband signals compared to baseline schemes without employing DAM. Moreover, it is shown that there exists a non-trivial trade-off between the adjustable RIS delay margin for aligning different reflected channels and the practical DAM component's power decay increased with RIS delay to achieve the maximum achievable rate.
\end{abstract}
\IEEEpeerreviewmaketitle
\section{Introduction}
\IEEEPARstart{T}{he} explosive growth of mobile data traffic have continuously driven the innovation of wireless communication technologies in the past decade, such as massive multiple-input multiple-output (MIMO) \cite{Larsson_CM_2014_Massive}, millimeter wave (mmWave) communications \cite{Rappaport_Access_2013_Millimeter}, and heterogeneous network (HetNet) \cite{Andrews_CM_2013_Seven}, as well as advanced channel coding design \cite{Bonello_CST_2011_low}. Looking forward to 2030 and beyond, the sixth-generation (6G) mobile networks are expected to provide 1,000 times higher network capacity (Terabits/second), 100 times higher connection density ($10^7$ devices/km$^2$), and 10 times lower latency (sub-millisecond) \cite{You_Science_2020_Towards}. Although massive MIMO and mmWave communications can achieve dramatic spectral efficiency improvements, the deployment of large-scale antenna arrays and the ever-increasing of bandwidth generally result in higher implementation cost and increased power consumption \cite{CISCO}. Additionally, it is hard to meet new communication requirements in the 6G vision by exploiting existing communication technologies, such as higher network energy efficiency, near 100\% coverage, higher mobility support, and higher intelligence resource allocation \cite{Saad_Network_2020_A}. As a result, how to develop green communications and enable intelligent connectivity for 6G mobile networks remains crucial \cite{Letaief_CM_2019_The}.\par
Recently, reconfigurable intelligent surface (RIS) and its various equivalents have been proposed as promising technologies to achieve the aforementioned goals \cite{Huang_TWC_2019_Reconfigurable, Wu_TWC_2019_Intelligent, Wu_CM_2020_Towards, Huang_WC_2020_Holographic, Di_JSAC_2020_Smart}. Specifically, RIS is a reconfigurable array comprising a huge number of passive reflecting elements, each of which is capable of inducing a phase shift and/or an attenuation to the incident signals and thereby customizing the desired wireless propagation environment according to the specific quality-of-service (QoS) requirements \cite{Wu_TC_2021_Intelligent, Wu_CM_2020_Towards, Dai_Access_2020_Reconfigurable}. By properly adjusting the phase shifts and attenuations caused by RIS elements, the reflected signals can be constructively superimposed with those from other paths at the desired receivers to enhance the received signal power and/or destructively to suppress the co-channel interference, both leading to improved wireless link performance \cite{Wu_TWC_2019_Intelligent, Huang_TWC_2019_Reconfigurable}. In contrast to the traditional amplify-and-forward/decode-and-forward relaying communication, RIS suffices to achieve higher energy efficiency in a full-duplex (FD) manner, without incurring the severe self-interference problem \cite{Bjornson_WCL_2020_Intelligent, Wu_CM_2020_Towards}. Hence, RIS is envisioned to become one of the potential solutions for enabling 6G mobile networks \cite{Zhang_JSAC_2020_Prospective}.\par
\begin{table*}[!t]
\tiny
\centering
\caption{A comparison of our contributions to existing work.}
\label{tab1}
\begin{tabular}{|l|l|l|c|c|c|l|c|}
\hline
\multicolumn{1}{|l|}{\multirow{2}{*}{\textbf{Schemes}}} & \multicolumn{1}{l|}{\multirow{2}{*}{\textbf{MIMO Setup}}} & \multicolumn{1}{l|}{\multirow{2}{*}{\textbf{Optimization Objective}}} & \multicolumn{1}{c|}{\multirow{2}{*}{\textbf{Frequency-selective}}} & \multicolumn{1}{c|}{\multirow{2}{*}{\textbf{Multiple RISs}}} & \multicolumn{3}{c|}{\textbf{RIS}}         \\ \cline{6-8} 
\multicolumn{1}{|c|}{}                           & \multicolumn{1}{c|}{}                            & \multicolumn{1}{c|}{}                                        & \multicolumn{1}{c|}{}                                             & \multicolumn{1}{c|}{}                               & \textbf{Attenuation} & \textbf{Phase Shift} & \textbf{Delay} \\ \hline
Huang \emph{et al.} \cite{Huang_TWC_2019_Reconfigurable}        &         Multi-user MISO           &      Energy efficiency/Sum-rate         &                \XSolidBrush           &        \XSolidBrush          &    \XSolidBrush         &     \cellcolor{mygray}\Checkmark Continuous       &  \XSolidBrush     \\ \hline
Wu \emph{et al.} \cite{Wu_TWC_2019_Intelligent}        &        Multi-user MISO            &   Transmit power      &         \XSolidBrush       &  \XSolidBrush          &      \XSolidBrush       &     \cellcolor{mygray}\Checkmark Continuous       &  \XSolidBrush     \\ \hline
Wu \emph{et al.} \cite{wu_TC_2020_Beamforming}        &        Multi-user MISO            &      Transmit power     &          \XSolidBrush      &          \XSolidBrush          &    \XSolidBrush         &      \cellcolor{mygray}\Checkmark Discrete      &  \XSolidBrush     \\ \hline
Han \emph{et al.} \cite{Han_TVT_2019_Large}        &          Single-user MISO          &     Ergodic spectral efficiency      &       \XSolidBrush      &       \XSolidBrush      &    \XSolidBrush      &  \cellcolor{mygray}\Checkmark Continuous      &  \XSolidBrush     \\ \hline
Cui \emph{et al.} \cite{Cui_WCL_2019_Secure}        &      Single-user MISO           &      Secrecy rate                                 &    \XSolidBrush          & \XSolidBrush  &      \XSolidBrush       &      \cellcolor{mygray}\Checkmark Continuous      &  \XSolidBrush     \\ \hline
Pan \emph{et al.} \cite{Pan_TWC_2020_Multicell}        &      Multi-cell multi-user MIMO             &    Weighted sum-rate       &      \XSolidBrush         &   \XSolidBrush          &    \XSolidBrush          &     \cellcolor{mygray}\Checkmark Continuous       &  \XSolidBrush     \\ \hline
Guo \emph{et al.} \cite{Guo_TWC_2020_Weighted}        &           Multi-user MISO         &     Weighted sum-rate                                  &               \XSolidBrush      &           \XSolidBrush          &     \XSolidBrush         &      \cellcolor{mygray}\Checkmark Continuous      &  \XSolidBrush     \\ \hline
Zhang \emph{et al.} \cite{Zhang_JSAC_2020_Capacity}        &       Single-user MIMO             &     Channel capacity                                  &                                        \cellcolor{mygray}\Checkmark        &                             \XSolidBrush          &    \XSolidBrush         &    \cellcolor{mygray}\Checkmark Continuous        &  \XSolidBrush     \\ \hline
Di \emph{et al.} \cite{Di_JSAC_2020_Hybrid}        &        Multi-user MISO            &       Sum-rate                                &             \XSolidBrush        &          \XSolidBrush          &    \cellcolor{mygray}\Checkmark         &        \cellcolor{mygray}\Checkmark Discrete    &  \XSolidBrush     \\ \hline
Abeywickrama \emph{et al.} \cite{Abeywickrama_TC_2020_Intelligent}        &      Multi-user MISO              &                 Transmit power                      &                                      \XSolidBrush          &                             \XSolidBrush          &             \multicolumn{2}{c|}{\cellcolor{mygray}\Checkmark Coupled}           &  \XSolidBrush     \\ \hline
Yang \emph{et al.} \cite{Yang_TC_2020_Intelligent}  &  Single-user SISO  &     Achievable rate    &    \cellcolor{mygray}\Checkmark    &                 \XSolidBrush                                    &      \cellcolor{mygray}\Checkmark       &     \cellcolor{mygray}\Checkmark Continuous       &   \XSolidBrush    \\ \hline
Zheng \emph{et al.} \cite{Zheng_WCL_2020_Intelligent}  & Single-user SISO  &   Achievable rate       &   \cellcolor{mygray}\Checkmark &                                        \XSolidBrush             &     \XSolidBrush        &     \cellcolor{mygray}\Checkmark Continuous       &    \XSolidBrush   \\ \hline
Our design      & Single-user SISO             &      Achievable rate        &  \cellcolor{mygray}\Checkmark    & \cellcolor{mygray}\Checkmark          &    \XSolidBrush       &       \cellcolor{mygray}\Checkmark Continuous     &   \cellcolor{mygray}\Checkmark    \\ \hline
\end{tabular}
	\vspace{-0.8cm}
\end{table*}
In RIS-assisted wireless systems, channel state information (CSI) acquisition and reflection pattern optimization constitute a pair of critical and fundamental problems \cite{An_TVT_2020_Optimal, Jensen_ICASSP_2020_An, You_JSAC_2020_Channel, Wang_TWC_2020_Channel, Huang_TWC_2019_Reconfigurable, Wu_TWC_2019_Intelligent}. For a completely passive RIS without installing any wireless transceiving modules, it is a blue moon to separately estimate the individual channels of the base station (BS)-RIS and RIS-user equipment (UE) links. In order to circumvent this issue, the authors of \cite{Jensen_ICASSP_2020_An, You_JSAC_2020_Channel, Wang_TWC_2020_Channel} proposed several novel methods estimating the cascaded BS-RIS-UE channels upon subtly designing estimation protocols and reflection patterns. Based on the estimated CSI, the reflection pattern optimization is then performed jointly with other design parameters for various optimization objectives, such as with the power allocation of the zero-forcing (ZF) precoding matrix for maximizing the energy efficiency \cite{Huang_TWC_2019_Reconfigurable}, with the transmit beamforming vectors for minimizing the transmit power \cite{Wu_TWC_2019_Intelligent, wu_TC_2020_Beamforming}, with the transmit beamforming vector for maximizing the secrecy rate \cite{Cui_WCL_2019_Secure, Shen_CL_2019_Secrecy, Yu_JSAC_2020_Robust}, and with the transmit precoding matrices of multiple BSs for maximizing the weighted sum rate \cite{Pan_TWC_2020_Multicell, Pan_JSAC_2020_Intelligent}. Moreover, in \cite{Han_TVT_2019_Large, Jung_TWC_2020_Performance}, only the statistical CSI is required for optimizing the reflection pattern since a large-scale reflection array generally results in a deterministic performance limit. More recently, a novel joint training method of the composite channel summing the direct and all reflected links was proposed to strike flexible trade-offs between the achievable rate performance and the pilot overhead \cite{An_WCL_2021_The, an_arxiv_2021_joint}. Additionally, the RIS reflection optimization has also been combined with other potential technologies, such as machine learning (ML) \cite{Huang_JSAC_2020_Reconfigurable, Elbir_WCL_2020_Deep}, unmanned aerial vehicles (UAV) \cite{Li_WCL_2020_Reconfigurable}, as well as mobile edge computing (MEC) \cite{Bai_TWC_2021_Resource}.\par
Nevertheless, it is worth noting that prior work on RIS-aided wireless systems have mostly considered frequency-flat fading channels for narrowband communications, where the RIS reflection coefficients are designed to align the phases of the reflected BS-RIS-UE links with that of the direct BS-UE link for coherent superposition. When frequency-selective fading channels are considered, the RIS reflection coefficients are required to cater to all subcarriers adamantly, for which the underlying optimization problem is thus more challenging to solve compared with the counterparts in narrow-band systems. To this end, in \cite{Zheng_WCL_2020_Intelligent, Yang_TC_2020_Intelligent, Lin_JSAC_2020_Adaptive, Zheng_TWC_2020_Intelligent}, the preliminary exploration is conducted upon applying RIS to orthogonal frequency division multiplexing (OFDM) communications. Specifically, the non-convex rate-maximizing problem was solved by alternating optimization (AO) algorithm relying on successive convex approximation (SCA) \cite{Yang_TC_2020_Intelligent} and a heuristic algorithm upon aligning the tap with the largest sum of channel gain \cite{Zheng_WCL_2020_Intelligent}. However, the strongest tap of different reflected channels might not be aligned due to the diverse time delays via different RIS elements, which chokes the performance improvement of RIS-assisted OFDM communications, especially for distributed RIS deployment.\par
Motivated by the above challenges, in this paper, we investigate RIS-enhanced OFDM communications over frequency-selective channels. In particular, we first introduce the delay adjustable metasurface (DAM), which is capable of adjusting the delays of signals reflected by different RIS elements \cite{Nakanishia_2018_APL_Storage}. Furthermore, we develop the rate-maximizing problem by jointly optimizing the power allocation and RIS reflection coefficients as well as RIS delays. The new degrees-of-freedom, i.e., delay, contributes to aligning the strongest taps of different reflected channels, thus improving the superimposed channel response in the frequency domain and resulting in a higher achievable rate. For the sake of illustration, we boldly and explicitly contrast our contributions to existing work in Table \ref{tab1}. Explicitly, the main contributions of this paper are summarized as follows:
\begin{itemize}
    \item Firstly, we introduce a novel DAM fabricated with varactor diodes \cite{Nakanishia_2018_APL_Storage}. In contrast to existing reflecting elements, each one in DAM is capable of storing and retrieving the impinging electromagnetic waves upon dynamically controlling its electromagnetically induced transparency (EIT) properties, thus additionally imposing an extra delay onto the incident signals.
    \item Secondly, we formulate an optimization problem aiming to maximize the achievable rate by jointly optimizing the power allocation at the transmitter and the reflection coefficients as well as the delays at the RIS. Nevertheless, the formulated problem is non-convex and thus non-trivial to solve optimally, for which we conceive an iterative algorithm that alternately optimizes the transmit power allocation and the RIS reflection pattern, including the reflection coefficients and the delays.
    \item Thirdly, in order to circumvent the high complexity for optimizing the RIS reflection coefficients, we propose a low-complexity scheme upon aligning the strongest taps of all reflected channels, while ensuring that the maximum delay spread after adding extra RIS delays does not exceed the length of the cyclic prefix (CP).
    \item Finally, numerical simulations evaluate the performance of our proposed designs. It is shown that the proposed scheme upon jointly optimizing the transmit power allocation and the RIS reflection pattern achieves better rate performance compared to the systems without employing DAM for both cases of perfect and estimated CSI. Furthermore, upon considering the power loss of the practical DAM component, it is unveiled that there generally exists a non-trivial trade-off between the delay and the decay that maximizes the achievable rate.
\end{itemize}\par
The rest of this paper is organized as follows. Section \ref{S2} introduces the system model of the point-to-point RIS-enhanced OFDM communications. Section \ref{S3} formulates the joint optimization problem of the OFDM power allocation and the RIS reflection pattern for maximizing the achievable rate. Furthermore, Section \ref{S4} proposes an AO algorithm to obtain a sub-optimal solution, while Section \ref{S5} conceives a low-complexity method and analyzes the theoretical performance. Section \ref{S6} provides numerical simulations to evaluate the performance of the proposed designs. Finally, Section \ref{S7} concludes the paper.\par
\emph{Notations:} Column vectors/matrices are denoted by bold-face lower/upper-case letters, while sets are indicated by upper-case calligraphic letters; For a matrix ${\bf{M}}$ of arbitrary size, ${{\bf{M}}^T}$, ${{\bf{M}}^*}$, and ${{\bf{M}}^H}$ represent the transpose, conjugate and Hermitian transpose, respectively; $\text{rank}\left( {\bf{M}} \right)$ returns the rank of matrix $\bf{M}$; while $\text{tr}\left( {\bf{V}} \right)$ represents the trace of square matrix ${\bf{V}}$; ${\bf{0}}_{x \times y}$ denotes an all-zero matrix of size ${x \times y}$; while ${\bf{1}}_{x \times y}$ denotes an all-one matrix of size ${x \times y}$; ${{\bf{I}}_M}$ represents the identity matrix of size $M$, while ${{\bf{e}}_m}$ denotes the $m$-th column of ${{\bf{I}}_M}$. Moreover, ${\text{diag}}\left( \bf{v} \right)$ denotes a diagonal matrix with the elements of $\bf{v}$ on its main diagonal and ${\text{diag}}\left( \bf{V} \right)$ represents a column vector formed by the main diagonals of square $\bf{V}$; $\left\| {\bf{v}} \right\|$ represents the Euclidean norm of vector ${\bf{v}}$; $\left|  v  \right|$ and $\angle v$ denote the modulus and the angle of a complex number $v$, respectively; while $\left\lfloor {a} \right\rceil $ represents the integer nearest to the real number $a$; $ \otimes $ denotes the Kronecker product and $ * $ stands for the linear convolution. Furthermore, $\mathbb{Z}^{x \times y}$, $\mathbb{R}^{x \times y}$, and $\mathbb{C}^{x \times y}$ denote the space of $x \times y$ integer-, real-, and complex-valued matrices, respectively; The $\log \left(  \cdot  \right)$ represents the logarithmic operation; while $\mathbb{E}\left(  \cdot  \right)$ stands for the expectation operation. The distribution of a circularly symmetric complex Gaussian (CSCG) random vector with mean vector ${\boldsymbol{\mu }}$ and covariance matrix ${\boldsymbol{\Sigma }}$ is denoted by ${\mathcal{CN}}\left( {{\boldsymbol{\mu }},{\boldsymbol{\Sigma }}} \right)$, while the distribution of a real-valued Gaussian random variable with mean ${\mu }$ and ${\sigma ^2}$ is denoted by $\mathcal{N}\left( {\mu ,{\sigma ^2}} \right)$; $\sim$ stands for ``distributed as''. $\max \left\{ {a,b} \right\}$ and $\min \left\{ {a,b} \right\}$ denote the maximum and minimum between two real numbers $a$ and $b$, respectively.
\begin{figure}[!t]
	\centering
	\includegraphics[width=8.8cm]{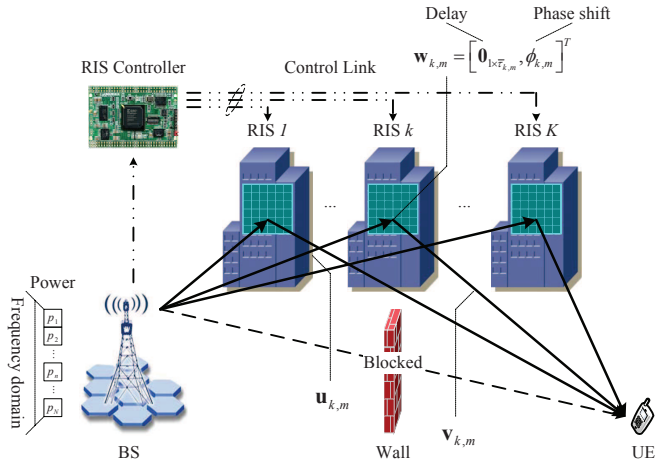}
	\caption{The considered multiple-RIS assisted OFDM communication system. RISs are assumed to be attached to surrounding buildings' facade.}
	\label{f1}
	\vspace{-0.8cm}
\end{figure}
\section{System Model}\label{S2}
As illustrated in Fig. \ref{f1}, we consider an OFDM wireless system, wherein multiple RISs are deployed on the surfaces of buildings or environmental objects to assist in the end-to-end transmission between a BS and a cell-edge UE. We assume that the direct BS-UE path is blocked by a wall. For the purpose of exposition, we assume that the BS and the UE are both equipped with a single antenna\footnote{In this paper, we consider the SISO scenario for the sake of characterizing the maximum performance gain benefited from delay adjustable RIS. When considering scenarios that BS is equipped with multiple antennas, the adjustable RIS delays have to cater to all channels spanning from different antennas to the user. The design problem considering adjustable RIS delay under MISO setups will be left as our future research topic.}. Explicitly, the total bandwidth allocated to the UE is equally divided into $N$ orthogonal subcarriers, which are denoted by the set ${\mathcal{N}} = \left\{ {0,1, \cdots ,N - 1} \right\}$. Furthermore, the number of RISs is assumed to be $K$, indexed by the set ${\mathcal{K}} = \left\{ {1,2, \cdots ,K} \right\}$. Each RIS is assumed to comprise $M$ passive reflecting elements\footnote{For the sake of illustration, we assume that all RISs are equipped with the same number of reflecting elements. For practical deployment, the number of RIS elements might be limited by the surface area of the attachment.}, denoted by the set ${\mathcal{M}} = \left\{ {1,2, \cdots ,M} \right\}$, and is connected to a smart RIS controller, which is capable of adjusting RIS reflection patterns in real-time for desired signal propagation \cite{Di_JSAC_2020_Smart, Wu_TC_2021_Intelligent}. In contrast to the existing RIS reflecting element, the one considered in this paper is also capable of imposing an extra time delay on the incident signals, which not only facilitate the coherent superposition of multiple copies of the desired signals but also maintain their synchronization in time. Specifically, one of the feasible approaches is to cascade the existing phase adjustable elements \cite{Wu_CM_2020_Towards} with the delay adjustable elements \cite{Nakanishia_2018_APL_Storage}. Moreover, we consider uplink channel training from the UE to the BS, where pilot symbols are sent from the UE and reflected by RISs, based on which the BS estimates all cascaded reflected BS-RIS-UE channels \cite{Zheng_WCL_2020_Intelligent, Yang_TC_2020_Intelligent}. For downlink communication, by assuming a time-division duplex (TDD) protocol and leveraging channel reciprocity, the reflection patterns can be readily optimized at the BS based on the CSI obtained during the uplink training and then informs all RISs via a separate wireless control link. Due to the severe path loss, it is further assumed that the power of the signals reflected by RISs more than once is negligible and thus are ignored, e.g., \cite{Wu_TWC_2019_Intelligent, Zhang_JSAC_2020_Capacity}.\par
In this paper, we consider a quasi-static frequency-selective block fading channel model for all channels involved in Fig. \ref{f1} and focus on one particular fading block where the channels remain approximately constant. Specifically, there exists an ${L_k}$-tap baseband equivalent multipath channel for the reflected BS-RIS-UE link via the $k$-th RIS\footnote{Note that here we have assumed that all the reflected channels via the reflecting elements at the same RIS experience the same cluster scatterers.}, through which the signals transmitted by the BS are reflected by the $k$-th RIS before arriving at the UE. Specifically, let ${{\bf{u}}_{k,m}} \in {\mathbb{C}^{{L_{k,\text{BR}}} \times 1}},\ \forall m \in {\mathcal{M}},\ \forall k \in {\mathcal{K}}$, denote the ${L_{k,\text{BR}}}$-tap baseband equivalent channel spanning from the BS to the $m$-th reflecting element at the $k$-th RIS. Similarly, let ${{\bf{v}}_{k,m}} \in {\mathbb{C}^{{L_{k,\text{RU}}} \times 1}},\ \forall m \in {\mathcal{M}},\ \forall k \in {\mathcal{K}}$, denote the ${L_{k,\text{RU}}}$-tap baseband equivalent channel of the RIS-UE link associated with the $m$-th reflecting element at the $k$-th RIS. Thus, we have ${L_k} = {L_{k,\text{BR}}} + {L_{k,\text{RU}}} - 1,\ \forall k \in {\mathcal{K}}$. Without loss of generality, we assume that ${L_1} \le {L_2} \le  \cdots  \le {L_K}$.\par
Furthermore, each RIS element is capable of rescattering the impinging signals upon imposing an independent reflection coefficient and a tunable time delay. Specifically, let ${{\boldsymbol{\phi}} _k} = {[ {{\phi _{k,1}},{\phi _{k,2}}, \cdots,}}$ ${{{\phi _{k,M}}} ]^T} \in {{\mathbb{C}}^{M \times 1}},\ \forall k \in {\mathcal{K}}$, denote the reflection coefficients at the $k$-th RIS, where each reflection coefficient ${\phi _{k,m}}$ characterizes the equivalent interaction of the $m$-th element on the incident signals at the $k$-th RIS. More explicitly, each ${\phi _{k,m}}$ can be expressed as\begin{small}\begin{align}
    {\phi _{k,m}} = {\beta _{k,m}}{e^{j{\theta _{k,m}}}},\ \forall m \in {\mathcal{M}},\ \forall k \in {\mathcal{K}},
\end{align}\end{small}where ${\beta _{k,m}} \in \left[ {0,1} \right]$ denotes the amplitude coefficient while ${\theta _{k,m}} \in \left[ {0,2\pi } \right)$ stands for the phase shift of the $m$-th element at the $k$-th RIS, respectively. To maximize the reflected power of the RIS and simplify its hardware design, we fix ${\beta _{k,m}} = 1,\ \forall m \in {\mathcal{M}},\ \forall k \in {\mathcal{K}}$, and only adjust the phase shift ${\theta _{k,m}}$ for reflection pattern optimization in this paper, thus we have $\left| {{\phi _{k,m}}} \right| = 1$. Furthermore, let ${{\boldsymbol{\tau}} _k} = {\left[ {{\tau _{k,1}},{\tau _{k,2}}, \cdots ,{\tau _{k,M}}} \right]^T} \in {\mathbb{R}^{M \times 1}},\ \forall k \in {\mathcal{K}}$, denote the delays at the $k$-th RIS, where each ${\tau _{k,m}}$ characterizes the delay imposed by the $m$-th element at the $k$-th RIS. For practical DAM operation, we have\begin{small}\begin{align}
  0 \le {\tau _{k,m}} \le {\tau _{\max }}, \ \forall m \in {\mathcal{M}}, \ \forall k \in {\mathcal{K}},
\end{align}\end{small}where ${\tau _{\max }}$ denotes the maximum time that a RIS element can store and retrieve the impinging signals \cite{Nakanishia_2018_APL_Storage}. For the sake of elaboration, we assume that the tunable delay resolution at each reflecting element is equal to the sampling interval\footnote{In fact, the adjustable delay resolution depends on the specific hardware design, which, anyhow, generally cause grid mismatch errors in time synchronization. Fortunately, with the employment of millimeter-wave and terahertz frequency bands, the mismatch errors will be substantially eliminated due to the increased sampling rate.}. Specifically, let ${\bar {\boldsymbol{\tau}} _k} = {\left[ {{{\bar \tau }_{k,1}},{{\bar \tau }_{k,2}}, \cdots ,{{\bar \tau }_{k,M}}} \right]^T} \in {{\mathbb{Z}}^{M \times 1}},\ \forall k \in {\mathcal{K}}$, denote the normalized delays at the $k$-RIS, where ${\bar \tau _{k,m}} = \left\lfloor {{\tau _{k,m}}{f_s}} \right\rceil,\ \forall m \in {\mathcal{M}}, \ \forall k \in {\mathcal{K}} $, denotes the discrete delay incurred by the $m$-th element at the $k$-th RIS, while ${f_s}$ represents the sampling rate. Thus, we have $0 \le {\bar \tau _{k,m}} \le {\bar \tau _{\max }},\ \forall m \in {\mathcal{M}},\ \forall k \in {\mathcal{K}}$, where ${\bar \tau _{\max }} = \left\lfloor {{\tau _{\max }}{f_s}} \right\rceil$ denotes the normalization of ${\tau _{\max }}$ with respect to ${f_s}$. As a result, the whole effect caused by the $m$-th reflecting element at the $k$-th RIS can be expressed by\begin{small}\begin{align}
    {{\bf{w}}_{k,m}} = {\left[ {{{\bf{0}}_{1 \times {{\bar \tau }_{k,m}}}},{\phi _{k,m}}} \right]^T},\ \forall m \in {\mathcal{M}},\ \forall k \in {\mathcal{K}},
\end{align}\end{small}which will degenerate into a conventional phase adjustable reflecting element if we have ${\bar \tau _{k,m}} = 0$. Moreover, note that here we do not take into account the power loss after experiencing the DAM in the resource allocation design. However, the power attenuation caused by DAM for practical implementation will be evaluated by our simulations in Section \ref{S6}.\par
Hence, the channel impulse response (CIR) of the BS-RIS-UE link reflected by the $m$-th reflecting element at the $k$-th RIS is thus the concatenation of the BS-RIS channel, RIS's rotation and delay, as well as the RIS-UE channel, which is given by ${{\bf{u}}_{k,m}} * {{\bf{w}}_{k,m}} * {{\bf{v}}_{k,m}} = {{\bf{w}}_{k,m}} * {{\bf{u}}_{k,m}} * {{\bf{v}}_{k,m}} \in {{\mathbb{C}}^{\left( {{{\bar \tau }_{k,m}} + {L_k}} \right) \times 1}},\ \forall m \in {\mathcal{M}},\ \forall k \in {\mathcal{K}}$. For the sake of exposition, define ${{\bf{H}}_k} = \left[ {{{\bf{h}}_{k,1}},{{\bf{h}}_{k,2}}, \cdots ,{{\bf{h}}_{k,M}}} \right] \in {{\mathbb{C}}^{N \times M}},\ \forall k \in {\mathcal{K}}$, as the zero-padded concatenated BS-RIS-UE channels without regard to the $k$-th RIS's effect, where we have ${{\bf{h}}_{k,m}} = {\left[ {{{\left( {{{\bf{u}}_{k,m}} * {{\bf{v}}_{k,m}}} \right)}^T},{{\bf{0}}_{1 \times \left( {N - {L_k}} \right)}}} \right]^T} \in {{\mathbb{C}}^{N \times 1}},\ \forall m \in {\mathcal{M}},\ \forall k \in {\mathcal{K}}$. The composite BS-RIS-UE channel of all $M$ reflected links via the $k$-th RIS, denoted by ${{\bf{g}}_k} \in {{\mathbb{C}}^{N \times 1}}$, can thus be expressed as\begin{small}\begin{align}
{{\bf{g}}_k} = \sum\limits_{m = 1}^M {{\phi _{k,m}}{{\bf{S}}_{k,m}}{{\bf{h}}_{k,m}}}  = {{\bf{S}}_k}{{\boldsymbol{\Phi }}_k}{{\bf{h}}_k},\ \forall k \in {\mathcal{K}},
\end{align}\end{small}where we have ${{\bf{S}}_{k,m}} = \left[ {{{\bf{0}}_{{{\bar \tau }_{k,m}} \times \left( {N - {{\bar \tau }_{k,m}}} \right)}},{{\bf{I}}_{{{\bar \tau }_{k,m}}}};{{\bf{I}}_{N - {{\bar \tau }_{k,m}}}},{{\bf{0}}_{\left( {N - {{\bar \tau }_{k,m}}} \right) \times {{\bar \tau }_{k,m}}}}} \right]$, ${{\bf{S}}_k} = \left[ {{{\bf{S}}_{k,1}},{{\bf{S}}_{k,2}}, \cdots ,{{\bf{S}}_{k,M}}} \right]$, ${{\bf{\Phi }}_k} = {\rm{diag}}\left( {{\phi _k}} \right) \otimes {{\bf{I}}_N}$, ${{\bf{h}}_k} = {\rm{vec}}\left( {{{\bf{H}}_k}} \right) = {\left[ {{\bf{h}}_{k,1}^T,{\bf{h}}_{k,2}^T, \cdots ,{\bf{h}}_{k,M}^T} \right]^T}$. We note that ${{\boldsymbol{\Phi }}_k}$ and ${{\bf{S}}_k}$ characterize the effect of reflection coefficients and delays of the $k$-th RIS, respectively. Therefore, the composite CIR from the BS to the UE is the superposition of all reflected BS-RIS-UE channels, which is given by\begin{small}\begin{align}
{\bf{\tilde g}} = \sum\limits_{k = 1}^K {{{\bf{g}}_k}}  = \sum\limits_{k = 1}^K {{{\bf{S}}_k}{{\boldsymbol{\Phi }}_k}{{\bf{h}}_k}} = \sum\limits_{k = 1}^K {\sum\limits_{m = 1}^M {{\phi _{k,m}}{{\bf{S}}_{k,m}}{{\bf{h}}_{k,m}}} }.
\end{align}\end{small}\par
Next, let us consider the downlink OFDM communications. Specifically, let ${\bf{p}} = {[ {{p_0},{p_1}, \cdots ,}}$ ${{{p_{N - 1}}} ]^T} \in {\mathbb{R}^{N \times 1}}$, where ${p_n} \ge 0,\ \forall n \in {\mathcal{N}}$, denotes the power allocated to the $n$-th subcarrier at the BS. Assume the aggregate transmit power available at the BS is $P$. Thus, the power allocation solution should satisfy $\sum\nolimits_{n = 0}^{N - 1} {{p_n}}  \le P$. Furthermore, let ${\bf{x}} = {\left[ {{x_0},{x_1}, \cdots ,{x_{N - 1}}} \right]^T}$ denote the normalized OFDM symbol, which is first transformed into the time domain via an $N$-point inverse discrete Fourier transform (IDFT), and then appended by a CP of length $N_{CP}$, which is assumed to be longer than the maximum delay spread of all reflected BS-RIS-UE channels, i.e., ${N_{CP}} \ge {L_K}$.\par
At the UE receiver, after removing CP and performing the $N$-point discrete Fourier transform (DFT), the equivalent baseband signal received in the frequency domain is given by\begin{small}\begin{align}
{\bf{y}} = {\bf{X}}{{\bf{P}}^{{1 \mathord{\left/
 {\vphantom {1 2}} \right.
 \kern-\nulldelimiterspace} 2}}}{{\bf{F}}}{\bf{\tilde g}} + {\bf{z}} = {\bf{X}}{{\bf{P}}^{{1 \mathord{\left/
 {\vphantom {1 2}} \right.
 \kern-\nulldelimiterspace} 2}}}{{\bf{F}}}\sum\limits_{k = 1}^K {\sum\limits_{m = 1}^M {{\phi _{k,m}}{{\bf{S}}_{k,m}}{{\bf{h}}_{k,m}}} }  + {\bf{z}},
\end{align}\end{small}where ${\bf{y}} = {\left[ {{y_0},{y_1}, \cdots ,{y_{N - 1}}} \right]^T}$ is the OFDM symbol received in the frequency domain; ${\bf{X}} = \text{diag}\left( {\bf{x}} \right)$ and ${\bf{P}} = \text{diag}\left( {\bf{p}} \right)$ denote the diagonal matrix of the transmit OFDM symbol ${\bf{x}}$ and the power allocation solution ${\bf{p}}$, respectively; ${\bf{z}} = {\left[ {{z_0},{z_1}, \cdots ,{z_{N - 1}}} \right]^T} \sim {\mathcal{CN}}\left( {{\bf{0}},{\sigma ^2}{{\bf{I}}_N}} \right)$ is the additive white Gaussian noise (AWGN) vector in the frequency domain with ${\sigma ^2}$ denoting the average noise power on each subcarrier; ${{\bf{F}}} \in {{\mathbb{C}}^{N \times N}}$ is the DFT matrix.\par
Specifically, the channel frequency response (CFR) at the $n$-th subcarrier is given by\begin{small}\begin{align}
    {d_n} = {\bf{f}}_n^H\sum\limits_{k = 1}^K {\sum\limits_{m = 1}^M {{\phi _{k,m}}{{\bf{S}}_{k,m}}{{\bf{h}}_{k,m}}} },\ \forall n \in {\mathcal{N}},
\end{align}\end{small}where ${\bf{f}}_n^H$ denotes the $n$-th row of the DFT matrix ${{\bf{F}}}$. Therefore, the achievable rate of RIS-enhanced OFDM systems in terms of bits per second per Hertz (b/s/Hz) is given by\begin{small}\begin{align}
R\left( {{\bf{p}},{\boldsymbol{\phi}},{\boldsymbol{\bar \tau }} } \right) = \frac{1}{{N + {N_{CP}}}}\sum\limits_{n = 0}^{N - 1} {{{\log }_2}\left( {1 + \frac{{{{\left| {{\bf{f}}_n^H\sum\limits_{k = 1}^K {\sum\limits_{m = 1}^M {{\phi _{k,m}}{{\bf{S}}_{k,m}}{{\bf{h}}_{k,m}}} } } \right|}^2}{p_n}}}{{\Gamma {\sigma ^2}}}} \right)}, \label{eq12}
\end{align}\end{small}where we have ${\boldsymbol{\phi}} {\rm{ = }}{\left[ {{\boldsymbol{\phi}} _1^T,{\boldsymbol{\phi}} _2^T, \cdots ,{\boldsymbol{\phi}} _K^T} \right]^T}$ and ${\boldsymbol{\bar \tau }}{\rm{ = }}{\left[ {{\boldsymbol{\bar \tau }}_1^T,{\boldsymbol{\bar \tau }}_2^T, \cdots ,{\boldsymbol{\bar \tau }}_K^T} \right]^T}$, while $\Gamma  \ge 1$ characterizes the gap from the Shannon's capacity owing to a practical modulation and coding scheme \cite{Goldsmith_book_2005_Wireless}.\par
\emph{Remark 1:} Note that in contrast to the conventional RIS-assisted OFDM systems that designs a common phase shift at each RIS element to cater to all subcarriers, e.g., \cite{Yang_TC_2020_Intelligent, Zheng_WCL_2020_Intelligent}, the new design degrees of freedom, ${{\bf{S}}_{k,m}}$, introduced in (\ref{eq12}) is capable of adjusting the RIS delay thus imposing differentiated impacts on the channel responses of different subcarriers. Therefore, the RIS-enhanced OFDM systems relying on DAM enables the coherent superposition of different reflected links on all subcarriers. ${{\bf{S}}_{k,m}}$ will be designed by our algorithms proposed in Sections \ref{S4} and \ref{S5}.\par
\emph{Remark 2:} Note that (\ref{eq12}) portrays the theoretical upper bound of the achievable rate of the considered RIS-enhanced OFDM systems, which is hardly achieved in practice. This is because to perform coherent detection at the receiver as well as to carry out the transmit power allocation and RIS reflection optimization, accurate knowledge of the CSI, i.e., $\left\{ {{{\bf{H}}_1},{{\bf{H}}_2}, \cdots, {{\bf{H}}_K}} \right\}$, is required, which needs to be acquired at the time and energy cost of extra channel training and feedback overhead. Additionally, the dimension of all reflected channels grows linearly with respect to the number of RISs, $K$, and the number of reflecting elements, $M$, which are practically very large and typically range from hundreds to thousands \cite{Tang_TWC_2021_Wireless, Dai_Access_2020_Reconfigurable}. Thus, the number of channel coefficients involved is much larger than that in conventional OFDM systems without employing RISs, which transparently scales up the required pilot overhead for training and results in higher complexity for CSI acquisition. Fortunately, since the RIS elements are generally closely packed, the channels associated with adjacent elements are practically correlated. Therefore, various grouping-based channel estimation approach have been proposed by arranging the adjacent RIS elements into a \emph{super-element} \cite{yu2021smart, Yang_TC_2020_Intelligent, You_JSAC_2020_Channel}, based on which the combined channel of each group is estimated and a common reflection coefficient in the same group is considered. For the sake of elaboration, we assume that the CSI of all reflected channels is a priori in the following, while the effects of channel estimation errors will be detailed by our simulations in Section \ref{S6}.
\section{Problem Formulation}\label{S3}
Given all reflected BS-RIS-UE channel matrices, i.e., $\left\{ {{{\bf{H}}_1},{{\bf{H}}_2}, \cdots ,{{\bf{H}}_K}} \right\}$, we aim to maximize the achievable rate shown in (\ref{eq12}) by jointly optimizing the transmit power allocation ${\bf{p}}$, the RIS reflection coefficients $\boldsymbol{\phi}$, as well as the RIS delays ${\boldsymbol{\bar \tau }}$. Therefore, we formulate the following optimization problem, where the constant terms in (\ref{eq12}) are omitted for brevity, that yields:\begin{small}\begin{subequations}\label{eq13}
\begin{alignat}{2}
{\left( {P1} \right)}: \ {\mathop {{\rm{max}}}\limits_{{\bf{p}},{\boldsymbol{\phi}},{\boldsymbol{\bar \tau}} } } \ &{\sum\limits_{n = 0}^{N - 1} {{{\log }_2}\left( {1 + \frac{{{{\left| {{\bf{f}}_n^H\sum\limits_{k = 1}^K {\sum\limits_{m = 1}^M {{\phi _{k,m}}{{\bf{S}}_{k,m}}{{\bf{h}}_{k,m}}} } } \right|}^2}{p_n}}}{{\Gamma {\sigma ^2}}}} \right)}} \label{13a}\\
{}{{\rm{s}}{\rm{.t}}{\rm{.}}} \ &{\sum\limits_{n = 0}^{N - 1} {{p_n}}  \le P}, \label{13b}\\
{}{} \ &{{p_n} \ge 0, \ \forall n \in {\mathcal{N}}}, \label{13c}\\
{}{} \ &{\left| {{\phi _{k,m}}} \right| = 1, \ \forall m \in {\mathcal{M}}, \ \forall k \in {\mathcal{K}}}, \label{13d}\\
{}{} \ &{0 \le {\bar \tau _{k,m}} \le {\bar \tau _{\max }}, \ \forall m \in {\mathcal{M}}, \ \forall k \in {\mathcal{K}}}, \label{13e}\\
{}{} \ &{{L_{k}} + {\bar \tau _{k,m}} \le {N_{CP}}, \ \forall m \in {\mathcal{M}}, \ \forall k \in {\mathcal{K}}}, \label{13f}
\end{alignat}
\end{subequations}\end{small}where (\ref{13b}) and (\ref{13c}) characterize the limitations on transmit power allocation; (\ref{13d}) denotes the constant-modulus RIS phase shift; (\ref{13e}) and (\ref{13f}) guarantee that the maximum delay spread after introducing extra RIS delays will not exceed the length of the CP. We note that Problem $\left( {P1} \right)$ is a non-convex optimization problem. Explicitly, it can be shown that the objective function (\ref{13a}) is non-concave over ${\boldsymbol{\phi}} $ and ${\boldsymbol{\bar \tau}} $; moreover, the variables ${\bf{p}}$, ${\boldsymbol{\phi}}$, and ${\boldsymbol{\bar \tau}}$ are coupled in the objective function (\ref{13a}), which makes their joint optimization difficult. To overcome the above challenges, in the following, we will propose an AO algorithm to find a high-quality sub-optimal solution for Problem $\left( {P1} \right)$, by iteratively optimizing one of ${\bf{p}}$ and $\left\{ {{\boldsymbol{\phi}} ,{\boldsymbol{\bar \tau}} } \right\}$ with the other fixed at each time.
\section{Joint Transmit Power Allocation and Reflection Pattern Optimization}\label{S4}
\subsection{Transmit Power Allocation Given RIS Reflection Pattern}\label{S4-1}
Note that given a set of RIS reflection pattern $\left\{ {{\boldsymbol{\phi}} ,{\boldsymbol{\bar \tau}} } \right\}$ as well as the CSI, the optimization problem in (\ref{eq13}) can be reduced to\begin{small}\begin{subequations}
\begin{alignat}{2}
{\left( {P2} \right)}: \ {\mathop {{\rm{max}}}\limits_{{\bf{p}} } } \ &{\sum\limits_{n = 0}^{N - 1} {{{\log }_2}\left( {1 + \frac{{{{\left| {{\bf{f}}_n^H\sum\limits_{k = 1}^K {\sum\limits_{m = 1}^M {{\phi _{k,m}}{{\bf{S}}_{k,m}}{{\bf{h}}_{k,m}}} } } \right|}^2}{p_n}}}{{\Gamma {\sigma ^2}}}} \right)}} \label{14a}\\
{}{{\rm{s}}{\rm{.t}}{\rm{.}}} \ &{\left( {\text{13a}} \right),\ \left( {\text{13b}} \right)}. \label{14b}
\end{alignat}
\end{subequations}\end{small}The optimal transmit power allocation ${\bf{p}}^o$ at the BS is thus given by the well-known water-filling (WF) solution \cite{Goldsmith_book_2005_Wireless}, i.e.,\begin{small}\begin{align}
{p_n^o} = {\left( {c - \frac{{\Gamma {\sigma ^2}}}{{{{\left| {{d_n}} \right|}^2}}}} \right)^ + }, \ \forall n \in {\mathcal{N}},\label{eq15}
\end{align}\end{small}where we have ${\left( a \right)^ + } \buildrel \Delta \over = \max \left( {0,a} \right)$, while ${c}$ is the cut-off power threshold that enables $\sum\nolimits_{n = 0}^{N - 1} {{p_n^o}}  = P$ that can be found by bisection search.
\subsection{RIS Reflection Pattern Optimization Given Transmit Power Allocation}
Given a transmit power allocation solution, \textbf{Problem $\left( {P1} \right)$} is thus simplified to\begin{small}\begin{subequations}
\begin{alignat}{2}
{\left( {P3} \right)}: \ {\mathop {{\rm{max}}}\limits_{{\boldsymbol{\phi}},{\boldsymbol{\bar \tau}} } } \ &{\sum\limits_{n = 0}^{N - 1} {{{\log }_2}\left( {1 + \frac{{{{\left| {{\bf{f}}_n^H\sum\limits_{k = 1}^K {\sum\limits_{m = 1}^M {{\phi_{k,m}}{{\bf{S}}_{k,m}}{{\bf{h}}_{k,m}}} } } \right|}^2}{p_n}}}{{\Gamma {\sigma ^2}}}} \right)}} \label{16a}\\
{}{{\rm{s}}{\rm{.t}}{\rm{.}}} \ &{\left( {\text{13d}} \right),\ \left( {\text{13e}} \right),\ \left( {\text{13f}} \right)}. \label{16b}
\end{alignat}
\end{subequations}\end{small}It can be shown that Problem $\left( {P3} \right)$ is still a non-convex optimization problem and thus non-trivial to maximize optimally. Alternatively, we consider maximizing the upper bound of (\ref{16a}) instead. Specifically, the achievable rate shown in (\ref{eq12}) is upper-bounded by\begin{small}\begin{align}\label{eq17}
R\left( {{\bf{p}},{\boldsymbol{\phi}} ,{\boldsymbol{\bar \tau }}} \right) \le \frac{N}{{N + {N_{CP}}}}{\log _2}\left( {1 + \frac{{\sum\limits_{n = 0}^{N - 1} {{{\left| {{\bf{f}}_n^H\sum\limits_{k = 1}^K {\sum\limits_{m = 1}^M {{\phi _{k,m}}{{\bf{S}}_{k,m}}{{\bf{h}}_{k,m}}} } } \right|}^2}{p_n}} }}{{N\Gamma {\sigma ^2}}}} \right),
\end{align}\end{small}based on Jensen's inequality. In fact, the upper bound of $R\left( {{\bf{p}},{\boldsymbol{\phi}} ,{\boldsymbol{\bar \tau }}} \right)$ on the right-hand-side (RHS) of (\ref{eq17}) is tight at high signal-to-noise-ratio (SNR) regions \cite{Goldsmith_book_2005_Wireless}, which is readily available under massive RIS deployment. Explicitly, Fig. \ref{f1_1} compares the achievable rate of (\ref{eq12}) and its upper bound of (\ref{eq17}), where we consider only a single RIS and the simple equal power allocation for the sake of illustration. Observed from Fig. \ref{f1_1} that the RHS of (\ref{eq17}) is an extremely tight upper bound of the achievable rate of (\ref{eq12}), especially for the RIS equipped with a large number of reflecting elements. Therefore, the maximization of the RHS of (\ref{eq17}) is almost equivalent to the original Problem $\left( {P3} \right)$.\par
After removing the constant terms, the new optimization problem is formulated as follows:\begin{small}\begin{subequations}
\begin{alignat}{2}
{\left( {P3{\text{-UB}}} \right)}: \ {\mathop {{\rm{max}}}\limits_{{\boldsymbol{\phi}},{\boldsymbol{\bar \tau}} } } \ &{\sum\limits_{n = 0}^{N - 1} {{{\left| {{\bf{f}}_n^H\sum\limits_{k = 1}^K {\sum\limits_{m = 1}^M {{\phi_{k,m}}{{\bf{S}}_{k,m}}{{\bf{h}}_{k,m}}} } } \right|}^2}{p_n}}} \label{18a}\\
{}{{\rm{s}}{\rm{.t}}{\rm{.}}} \ &{\left( {\text{13d}} \right),\ \left( {\text{13e}} \right),\ \left( {\text{13f}} \right)}, \label{18b}
\end{alignat}
\end{subequations}\end{small}which turns out to be the maximization of the weighted-sum of CFRs at the receiver.\par
Furthermore, note that for a given tentative delay vector ${\boldsymbol{\bar \tau}} $, Problem $\left( {P3\text{-UB}} \right)$ reduces to \begin{table}[!t]
    \small
	\centering
	\caption{\label{tab3}} 
	\begin{tabular}{p{15cm}}
		\toprule
		{\bfseries Algorithm 1:} Alternating Optimization (AO) Algorithm for Solving Problem $\left( {P1} \right)$\\
		\midrule
	1: {\bfseries Input:} $\left\{ {{{\bf{H}}_1},{{\bf{H}}_2}, \cdots ,{{\bf{H}}_K}} \right\}$, $\Gamma $, ${\sigma ^2}$.\\
	2: Randomly generate multiple initializations, i.e., $\left\{ {{{\bf{p}}^{\left( 1 \right)}},{{\bf{p}}^{\left( 2 \right)}}, \cdots ,{{\bf{p}}^{\left( J \right)}}} \right\}$; \\  
	3: {\bfseries for} ${{\bf{p}}^{\left( j \right)}},\ 1 \le j \le J$, \bf{do}  \\ 
	4: \quad {\bfseries for} ${\boldsymbol{\bar \tau }} \in {\mathcal{A}}$, {\bfseries do}\\  
	5: \qquad Solve Problem $\left( {P4\text{-SDR}} \right)$ based on the given $\bf{p}$ and ${\boldsymbol{\bar \tau }}$ via CVX;  \\
		6: \qquad Apply Gaussian randomization to find an approximate ${\boldsymbol{\phi}}$ of Problem $\left( {P4\text{-E}} \right)$;\\
	7: \quad{\bfseries end for}\\
	8: \quad Solve Problem $\left( {P5} \right)$ to select the optimal ${\boldsymbol{\bar \tau }}$ and ${\boldsymbol{\phi}}$;\\
	9: \quad Solve Problem $\left( {P2} \right)$ to obtain the optimal ${\bf{p}}$ based on $\left\{ {{\boldsymbol{\phi}}, {\boldsymbol{\bar \tau }} } \right\}$; \\
	10: {\bf{while}} The objective value of (\ref{13a}) with the obtained ${\bf{p}}^o$, ${\boldsymbol{\phi}}^o$ and ${\boldsymbol{\bar \tau }}^o$ reaches convergence;\\
	11: {\bf{Output:}} The optimal $\left\{ {{\bf{p}}^o, {\boldsymbol{\phi}}^o, {\boldsymbol{\bar \tau }}^o} \right\}$ maximizing the achievable rate of (\ref{eq12}) from $J$ candidates.\\
		\bottomrule
	\end{tabular}
		\vspace{-0.8cm}
\end{table}
\begin{small}\begin{subequations}
\begin{alignat}{2}
{\left( {P4} \right)}: \ {\mathop {{\rm{max}}}\limits_{{\boldsymbol{\phi}} } } \ &{\sum\limits_{n = 0}^{N - 1} {{{\left| {{\bf{f}}_n^H\sum\limits_{k = 1}^K {\sum\limits_{m = 1}^M {{\phi_{k,m}}{{\bf{S}}_{k,m}}{{\bf{h}}_{k,m}}} } } \right|}^2}{p_n}}} \label{19a}\\
{}{{\rm{s}}{\rm{.t}}{\rm{.}}} \ &{\left( {\text{13d}} \right)}, \label{19b}
\end{alignat}
\end{subequations}\end{small}which can be further simplified to
\begin{small}\begin{subequations}
\begin{alignat}{2}
{\left( {P4\text{-E}} \right)}: \ {\mathop {{\rm{max}}}\limits_{{\boldsymbol{\phi}} } } \ &{{\left\| {{{\bf{P}}^{{1 \mathord{\left/
 {\vphantom {1 2}} \right.
 \kern-\nulldelimiterspace} 2}}}{\bf{FT}}{\boldsymbol{\phi}}  } \right\|^2}} \label{20a}\\
{}{{\rm{s}}{\rm{.t}}{\rm{.}}} \ &{\left( {\text{13d}} \right)}, \label{20b}
\end{alignat}
\end{subequations}\end{small}where ${\bf{T}}$ is defined by ${\bf{T}} = \left[ {{{\bf{S}}_{1,1}}{{\bf{h}}_{1,1}}, \cdots ,{{\bf{S}}_{1,M}}{{\bf{h}}_{1,M}}, \cdots ,{{\bf{S}}_{K,1}}{{\bf{h}}_{K,1}}, \cdots ,{{\bf{S}}_{K,M}}{{\bf{h}}_{K,M}}} \right]$. Note that Problem $\left( {P4\text{-E}} \right)$ is a non-convex quadratically constrained quadratic problem (QCQP), for which we can apply the semidefinite relaxation (SDR) \cite{Wu_TWC_2019_Intelligent, Yang_TC_2020_Intelligent} technique to obtain an approximate solution. Specifically, define ${\boldsymbol{\Psi }} = {\boldsymbol{\phi}} {{\boldsymbol{\phi}} ^H}$ and ${\bf{R}} = {{\bf{T}}^H}{{\bf{F}}^H}{\bf{PFT}}$, we transform Problem $\left( {P4\text{-E}} \right)$ into the following problem by applying SDR to relax the rank-one constraint, i.e.,\begin{small}\begin{subequations}
\begin{alignat}{2}
{\left( {P4\text{-SDR}} \right)}: \ {\mathop {{\rm{max}}}\limits_{{\boldsymbol{\phi}} } } \ &{{\text{tr}}\left( { {\bf{R}}{\boldsymbol{\Psi }} } \right)} \label{22a}\\
{}{{\rm{s}}{\rm{.t}}{\rm{.}}} \ &{{{\bf{\Psi }}_{i,i}} = 1,\ 1 \le i \le MK}, \label{22b}\\
{}{} \ &{{{\bf{\Psi }}} \succeq {\bf{0}}}. \label{22c}
\end{alignat}
\end{subequations}\end{small}Problem $\left( {P4\text{-SDR}} \right)$ is a convex semidefinite programming (SDP) problem, which can be efficiently solved via existing convex optimization software, e.g., CVX \cite{cvx}. Let ${{\bf{\Psi }}^o}$ denote the optimal solution to Problem $\left( {P4\text{-SDR}} \right)$. If we have ${\text{rank}}\left( {{{\bf{\Psi }}^o}} \right) = 1$, the relaxation from Problem $\left( {P4\text{-E}} \right)$ to Problem $\left( {P4\text{-SDR}} \right)$ is tight and the optimal $\boldsymbol{\phi}$ to Problem $\left( {P4\text{-E}} \right)$ can be obtained by ${\boldsymbol{\phi}}  = {\bf{U}}\text{diag}\left( {{{\bf{\Lambda }}^{{1 \mathord{\left/
 {\vphantom {1 2}} \right.
 \kern-\nulldelimiterspace} 2}}}} \right)$, where ${{\bf{\Psi }}^o} = {\bf{U\Lambda }}{{\bf{U}}^H}$ is the eigen-value decomposition (EVD) of the matrix ${{\bf{\Psi }}^o}$. By contrast, if ${\rm{rank}}\left( {{{\bf{\Psi }}^o}} \right) > 1$ holds, the optimal objective value of Problem $\left( {P4\text{-SDR}} \right)$ serves as an upper bound to that of Problem $\left( {P4\text{-E}} \right)$ and thus we have to construct a rank-one solution according to ${{\bf{\Psi }}^o}$. In this paper, we consider a customized Gaussian randomization method \cite{luo_SPM_2010_Semidefinite} to find an approximate solution to Problem $\left( {P4\text{-E}} \right)$. Specifically, a number (denoted by $Q$) of $\boldsymbol{\phi}$ are generated by ${{\boldsymbol{\phi}} _q} = {\bf{U}}{\text{diag}}\left( {{{\bf{\Lambda }}^{{1 \mathord{\left/
 {\vphantom {1 2}} \right.
 \kern-\nulldelimiterspace} 2}}}} \right){\boldsymbol{\delta }},\ q = 1,2, \cdots ,Q$, where ${\boldsymbol{\delta }} \sim {\mathcal{CN}}\left( {{\bf{0}},{{\bf{I}}_{KM}}} \right)$ is a complex-valued Gaussian random vector. Following this, a sub-optimal reflection coefficient solution to Problem $\left( {P4\text{-E}} \right)$ is thus obtained by selecting the one from $\left\{ {{{\boldsymbol{\phi}} _1},{{\boldsymbol{\phi}} _2}, \cdots ,{{\boldsymbol{\phi}} _Q}} \right\}$ that maximizes the objective value of (\ref{20a}).\par
For each legitimate ${\boldsymbol{\bar \tau}} $, we repeat to solve the formulated Problem $\left( {P4\text{-E}} \right)$. After obtaining the sub-optimal ${\boldsymbol{\phi}}$ for all feasible ${\boldsymbol{\bar \tau}} $, the optimal solution of ${\boldsymbol{\bar \tau}}$ can be readily obtained by solving\begin{small}\begin{subequations}
\begin{alignat}{2}
{\left( {P5} \right)}: \ {\mathop {{\rm{max}}}\limits_{{\boldsymbol{\bar \tau}} } } \ &{{\left\| {{{\bf{P}}^{{1 \mathord{\left/
 {\vphantom {1 2}} \right.
 \kern-\nulldelimiterspace} 2}}}{\bf{FT}}\left( {{\boldsymbol{\bar \tau }}} \right){\boldsymbol{\phi}} \left( {{\boldsymbol{\bar \tau }}} \right)} \right\|^2}} \label{23a}\\
{}{{\rm{s}}{\rm{.t}}{\rm{.}}} \ &{0 \le {{\bar \tau }_{k,m}} \le {{\bar \tau }'_k},\ \forall m \in {\mathcal{M}},\ \forall k \in {\mathcal{K}}}, \label{23b}
\end{alignat}
\end{subequations}\end{small}where we have ${{\bar \tau }'_k} = \min \left( {{{\bar \tau }_{\max }},{N_{CP}} - {L_k}} \right)$. We note that Problem $\left( {P5} \right)$ can be readily solved by searching over all possible legitimate delay vectors. Since there is no much delay difference between the different reflected channels via the same RIS with moderate size, we can reduce the complexity of solving Problem $\left( {P5} \right)$ upon searching for a common delay for each RIS. The relationship between the tolerable delay and its corresponding RIS size is detailed in Section \ref{S5-2}. For each RIS, the scope of the adjustable delay does not exceed the CP margin. Let ${{\boldsymbol{\bar \tau }}^o}$ denote the optimal solution to Problem $\left( {P5} \right)$, thus the high-quality sub-optimal solution to Problem $\left( {P3\text{-UB}} \right)$ is given by $\left\{ {{{{\boldsymbol{\bar \tau }}}^o},{\boldsymbol{\phi}} \left( {{{{\boldsymbol{\bar \tau }}}^o}} \right)} \right\}$.\par
In summary, the overall AO algorithm for solving Problem $\left( {P1} \right)$ is given in \textbf{Algorithm 1}, where $\mathcal{A}$ in line $4$ is defined by ${\mathcal{A}} = {\left[ {0,{{\bar \tau }'_1}} \right]^M} \times {\left[ {0,{{\bar \tau }'_2}} \right]^M} \times  \cdots  \times {\left[ {0,{{\bar \tau }'_K}} \right]^M}$. We note that AO algorithm might fall into a local optimal solution. Hence, in \textbf{Algorithm 1}, we provide a heuristic method to address this problem upon adopting multiple (denoted by $J$) random initializations \cite{Zhang_JSAC_2020_Capacity}, which guarantees that \textbf{Algorithm 1} finds a high-quality sub-optimal solution to the original Problem $\left( {P1} \right)$ at least. Note that the complexity of \textbf{Algorithm 1} is critically high. In the next section, we will propose a low-complexity method for solving Problem $\left( {P1} \right)$ efficiently.
\section{A Low-Complexity Optimization Method for Power Allocation and Reflection Optimization}\label{S5}
\subsection{The Proposed Low-Complexity Method}
Although the AO algorithm in Table \ref{tab3} achieves high-quality sub-optimal performance, its complexity for solving Problem $\left( {P1} \right)$ is shown in the order of ${\mathcal{O}}\left( {{N_{i}}{K^{4.5}}{M^{4.5}}\prod\nolimits_{k = 1}^K {{\bar \tau' _k}} } \right)$, where ${N_{i}}$ represents the number of iteration times, ${\mathcal{O}}\left( {{K^{4.5}}{M^{4.5}}} \right)$ characterizes the complexity order for solving Problem $\left( {P4\text{-SDR}} \right)$ at each iteration \cite{Zheng_WCL_2020_Intelligent, luo_SPM_2010_Semidefinite}, while ${{\bar \tau' _k}}$ is the cardinality of the feasible delay set associated with the $k$-th RIS. We note that the complexity of \textbf{Algorithm 1} is practically costly for large values of $K$ and $M$. Hence, we propose in this subsection a low-complexity method to solve Problem $\left( {P1} \right)$ suboptimally by focusing on the delay domain. Specifically, for a typical wireless channel, we have ${L_{K}} \le {N_{CP}} \ll N$, which implies that the channel gain is much more concentrated in the delay domain than that in the frequency domain. Therefore, upon leveraging Parseval's theorem and omitting the transmit power allocation, the objective function of (\ref{18a}) can be transformed into the delay domain as\begin{small}\begin{subequations}
\begin{alignat}{2}
{\left( {P6} \right)}: \ {\mathop {{\rm{max}}}\limits_{{\boldsymbol{\phi}},{\boldsymbol{\bar \tau}} } } \ &{\sum\limits_{n = 0}^{N - 1} {{{\left| {\sum\limits_{k = 1}^K {\sum\limits_{m = 1}^M {{\phi_{k,m}}{{\bf{S}}_{k,m}}{{\bf{h}}_{k,m}}} } } \right|}^2}}} \label{24a}\\
{}{{\rm{s}}{\rm{.t}}{\rm{.}}} \ &{\left( {\text{13d}} \right), \ \left( {\text{13e}} \right), \ \left( {\text{13f}} \right)}. \label{24b}
\end{alignat}
\end{subequations}\end{small}\par
Next, we propose our low-complexity method by adjusting RISs' reflection coefficients and delays to align the strongest tap of all reflected channels, while ensuring that the maximum delay spread after adding extra delay does not exceed the length of the CP. Specifically, we first find the tap, denoted by ${\hat l_{k,m}},\ \forall m \in \mathcal{M},\ \forall k \in \mathcal{K}$, having the largest CIR gain with respect to the $m$-th element at the $k$-th RIS, i.e.,\begin{small}\begin{align}\label{eq25}
{\hat l_{k,m}} = \arg \mathop {\max }\limits_{{l_{k,m}} \in \left\{ {1,2, \cdots ,{L_k}} \right\}} {\left| {{{\bf{h}}_{k,m}}\left( {l_{k,m}} \right)} \right|^2}.
\end{align}\end{small}Furthermore, in order to eliminate the inter-symbol interference, we have to ensure that the maximum difference of indices found by (\ref{eq25}) is within the reasonable range of the adjustable delay. To achieve this, we reseek the strongest tap within a feasible tap scope determined by the maximum index value of ${\hat l_{\max }} = \mathop {\max }\limits_{m \in \mathcal{M},k \in \mathcal{K}} \left\{ {{{\hat l}_{k,m}}} \right\}$, i.e.,\begin{small}\begin{align}\label{eq26}
{\tilde l_{k,m}} = \arg \mathop {\max }\limits_{{l_{k,m}} \in {{\mathcal{L}}_k}} {\left| {{{\bf{h}}_{k,m}}\left( {{l_{k,m}}} \right)} \right|^2},\ \forall m \in \mathcal{M},\ \forall k \in \mathcal{K},
\end{align}\end{small}where we have ${{\mathcal{L}}_k} = \left\{ {{{\hat l}_{\max }} - {{\bar \tau }'_k},{{\hat l}_{\max }} - {{\bar \tau }'_k} + 1, \cdots ,{{\hat l}_{\max }}} \right\}$. Hence, the suboptimal reflection coefficients ${\phi _{k,m}^{o}}$ and delays ${\bar \tau _{k,m}^{o}}$, $\forall m \in \mathcal{M},\ \forall k \in {\mathcal{K}}$, for aligning the strongest tap are given by\begin{small}\begin{align}\label{eq27}
    {\phi _{k,m}^{o}} = {e^{ - j\angle {{\bf{h}}_{k,m}}\left( {{{\tilde l}_{k,m}}} \right)}},\ {\bar \tau _{k,m}^{o}} = {\tilde l_{\max }} - {\tilde l_{k,m}},
\end{align}\end{small}respectively, where we have ${\tilde l_{\max }} = \mathop {\max }\limits_{m \in \mathcal{M},k \in \mathcal{K}} \left\{ {{{\tilde l}_{k,m}}} \right\}$.\par
After obtaining the RISs' reflection coefficients and delays, the optimal WF solution is invoked to perform the transmit power allocation, as described in Section \ref{S4-1}. The detailed procedure of this low-complexity method is summarized in Table \ref{tab4}, which is referred to as the strongest tap alignment (STA) method. The complexity of the STA method is shown in the order of ${\mathcal{O}}\left( {M\sum\nolimits_{k = 1}^K {{L_k}} } \right)$, which is significantly reduced compared to that of \textbf{Algorithm 1}. It is worth pointing out that \cite{Zheng_WCL_2020_Intelligent} introduced the method of aligning the taps of all reflected channels resulting in the largest sum of CIR gain in the absence of DAM, namely, adjusting only the RIS reflection coefficients via (\ref{eq27}) upon substituting the index value of $\tilde l = \arg \mathop {\max }\limits_{l \in \left\{ {0,1, \cdots ,{L_K}} \right\}} {\left| {\sum\nolimits_{k = 1}^K {\sum\nolimits_{m = 1}^M {\left| {{{\bf{h}}_{k,m}}\left( l \right)} \right|} } } \right|^2}$ for $\forall m \in \mathcal{M},\ \forall k \in \mathcal{K}$, which obviously severs as a lower bound of our STA method due to the fact that the sum of the maximum is greater than the maximum of the sum. Apparently, the improved performance gain is due to the new design degrees of freedom, i.e., delay.
\begin{table}[!t]
    \small
	\centering
	\caption{\label{tab4}} 
	\begin{tabular}{p{15cm}}
		\toprule
		{\bfseries Algorithm 2:} The Low-Complexity Strongest Tap Alignment (STA) Method for Solving Problem $\left( {P1} \right)$\\
		\midrule
		1: {\bfseries Input:}       $\left\{ {{{\bf{H}}_1},{{\bf{H}}_2}, \cdots ,{{\bf{H}}_K}} \right\}$, $\Gamma $, ${\sigma ^2}$.\\
		2: Find the strongest tap for each reflected channel by (\ref{eq25});  \\
		3: Update the strongest tap into a feasible scope by (\ref{eq26});  \\
		4: Calculate the RISs' coefficients ${\boldsymbol{\phi}}^o$ and delays ${\boldsymbol{\bar \tau }}^o$ by (\ref{eq27});  \\ 
		5: Perform the WF algorithm to obtain ${\bf{p}}^o$;  \\  
		6: {\bf{Output:}} $\left\{ {{\bf{p}}^o,{\boldsymbol{\phi}}^o ,{\boldsymbol{\bar \tau }}^o} \right\}$.\\
		\bottomrule
	\end{tabular}
	\vspace{-0.8cm}
\end{table}
\subsection{Theoretical Analysis}\label{S5-2}
Next, we will provide some theoretical analysis to verify the benefits of DAM. Specifically, the effects of DAM upon considering all reflected channels having only a non-zero tap are summarized in \textbf{\emph{Lemma 1}}, where only a single reflecting element is mounted at each RIS for the sake of illustration.\par
\textbf{\emph{Lemma 1:}} Let ${{\bf{h}}_k} = \left| {{h_k}} \right|{e^{j\angle {h_k}}} {{\bf{e}}_{{L _k}}} \in {{\mathbb{C}}^{N \times 1}},\ k \in \mathcal{K}$, denote the CIR of the $k$-th reflected BS-RIS-UE channel. Hence, DAM is capable of ensuring the perfect coherent superposition of $K$ reflected copies on all subcarriers. Explicitly, the optimal reflection coefficient and delay for the $k$-th RIS element are given by ${\phi _k} = {e^{ - j\angle {h_k}}}, \ {\bar \tau _k} = {L _K} - {L_k},\ k \in \mathcal{K}$, respectively.\par
\textbf{\emph{Proof:}} Without loss of generality, we assume ${L _1} \le {L _2} \le  \cdots  \le {L _K}$. The CFR of the $k$-th reflected channel at the $n$-th subcarrier is given by
\begin{small}\begin{align}
    {d_{n,k}} = {\bf{f}}_n^H{{\bf{h}}_k} = {\bf{f}}_n^H\left| {{h_k}} \right|{e^{j\angle {h_k}}}{{\bf{e}}_{{L _k}}},\ \forall k \in \mathcal{K},\ \forall n \in \mathcal{N}.
\end{align}\end{small}In order to facilitate the coherent superposition of the CFR of the $k$-th channel and the $K$-th channel on the $n$-th subcarrier, the reflection coefficient at the $k$-th RIS for catering to the $n$-th subcarrier, denoted by ${\kappa _{n,k}},\ \forall k \in \mathcal{K},\ \forall n \in \mathcal{N}$, can be expressed as\begin{small}\begin{align}
{\kappa _{n,k}} = \frac{{{{{d_{n,K}}} \mathord{\left/
 {\vphantom {{{d_{n,K}}} {{d_{n,k}}}}} \right.
 \kern-\nulldelimiterspace} {{d_{n,k}}}}}}{{\left| {{{{d_{n,K}}} \mathord{\left/
 {\vphantom {{{d_{n,K}}} {{d_{n,k}}}}} \right.
 \kern-\nulldelimiterspace} {{d_{n,k}}}}} \right|}} = {e^{j\angle {h_K}}}{e^{ - j\angle {h_k}}}{\bf{f}}_n^H{{\bf{e}}_{{L_K}}}{\bf{e}}_{{L_k}}^H{{\bf{f}}_n}.
\end{align}\end{small}It can be readily seen that if and only if ${{\bf{e}}_{{L _k}}} = {{\bf{e}}_{{L_K}}}$, ${\kappa _{n,k}}$ is independent of $n$. Thus we have ${\phi _k} = {e^{ - j\angle {h_k}}},\ {\bar \tau _k} = {L _K} - {L _k},\  k \in \mathcal{K}$. Since we have ${L_K} \le {N_{CP}}$, thus the lengths of all CIRs after adding extra delays will not exceed the length of the CP. The proof is completed. $\hfill\blacksquare$\par
\textbf{\emph{Lemma 1}} demonstrates that for one-tap reflected channels, DAM is capable of facilitating perfect coherent superposition of all reflected channels on all subcarriers. Although the one-tap assumption in \textbf{\emph{Lemma 1}} is rigorous, in practical systems, the BS-RIS and RIS-UE channels are generally of line-of-sight (LoS) propagation, with a large proportion of power in the LoS path \cite{Wu_TWC_2019_Intelligent, Di_JSAC_2020_Smart}. In this sense, the role of DAM is to align these LoS components of different reflected channels, especially for the distributed RIS deployment resulting in distinctly different delays.\par
Furthermore, even if only a single RIS is adopted, the delays via different RIS elements are no longer negligible if the RIS's size is large enough. Specifically, considering a square RIS array employing $M$ elements with element spacing of $d$, the RIS size incurring negligible delay should satisfies $2\sqrt 2 \left( {\sqrt M  - 1} \right)d < {c \mathord{\left/
 {\vphantom {c {2{f_s}}}} \right.
 \kern-\nulldelimiterspace} {2{f_s}}}$, where $c = 3.0 \times {10^8}$ m/s is the velocity of light. Therefore, we have $M < {\left( {{c \mathord{\left/
 {\vphantom {c {4\sqrt 2 {f_s}d}}} \right.
 \kern-\nulldelimiterspace} {4\sqrt 2 {f_s}d}} + 1} \right)^2} \approx {{{c^2}} \mathord{\left/
 {\vphantom {{{c^2}} {32f_s^2{d^2}}}} \right.
 \kern-\nulldelimiterspace} {32f_s^2{d^2}}}$ characterizing the largest size of RIS resulting in negligible delay effects. Given an example considering the carrier frequency of ${f_c} = 3$ GHz and the sampling rate of ${f_s} = 50$ MHz, thus the maximum number of RIS elements with a tolerable time delay is ${M_{\max }} = 450$ for the RIS equipped with half-wavelength spaced elements.\par
\begin{figure}[!t]
	\centering
	\subfigure[\label{f1_1}]{\includegraphics[width=5.4cm]{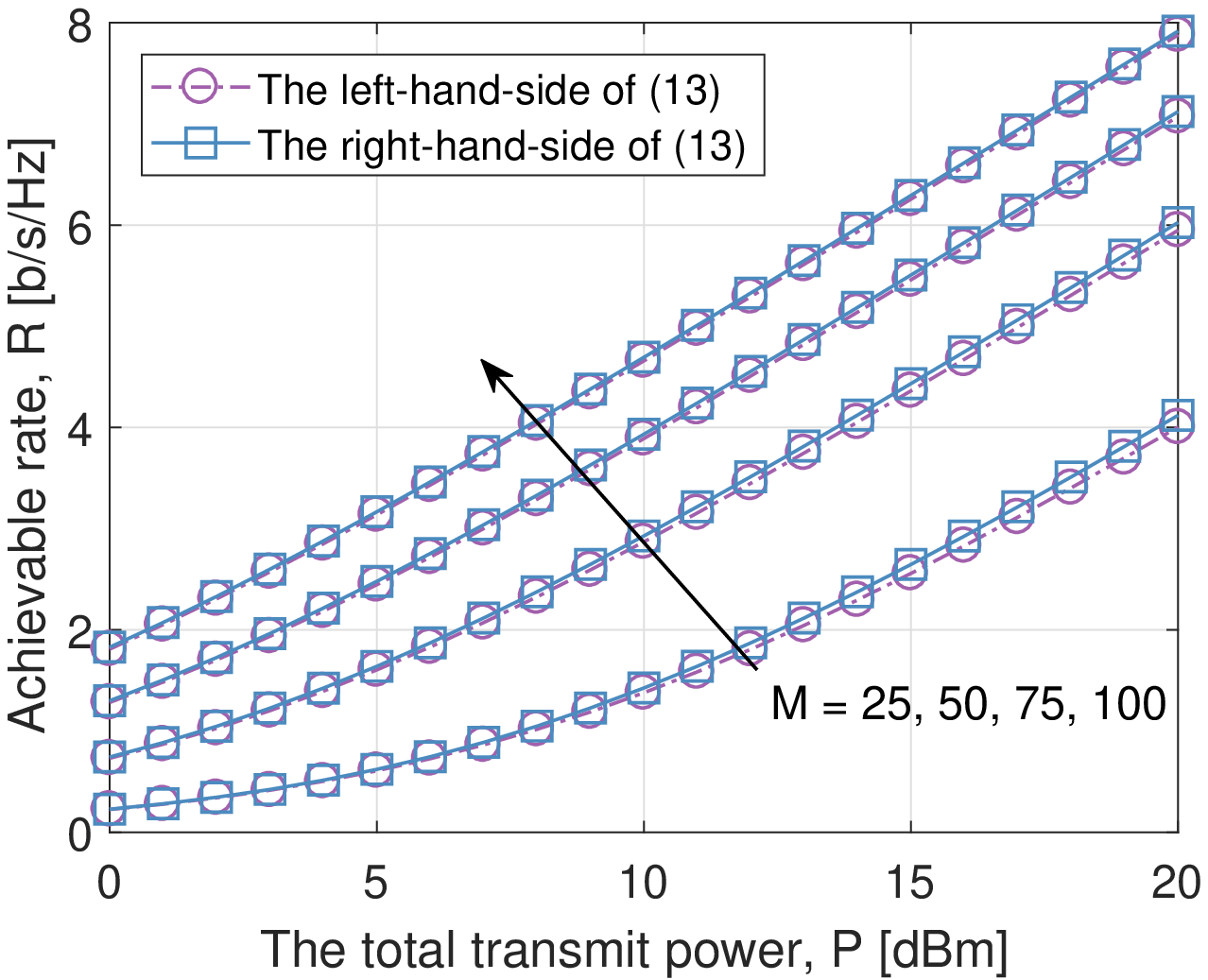}}
	\subfigure[\label{f2}]{\includegraphics[width=5.4cm]{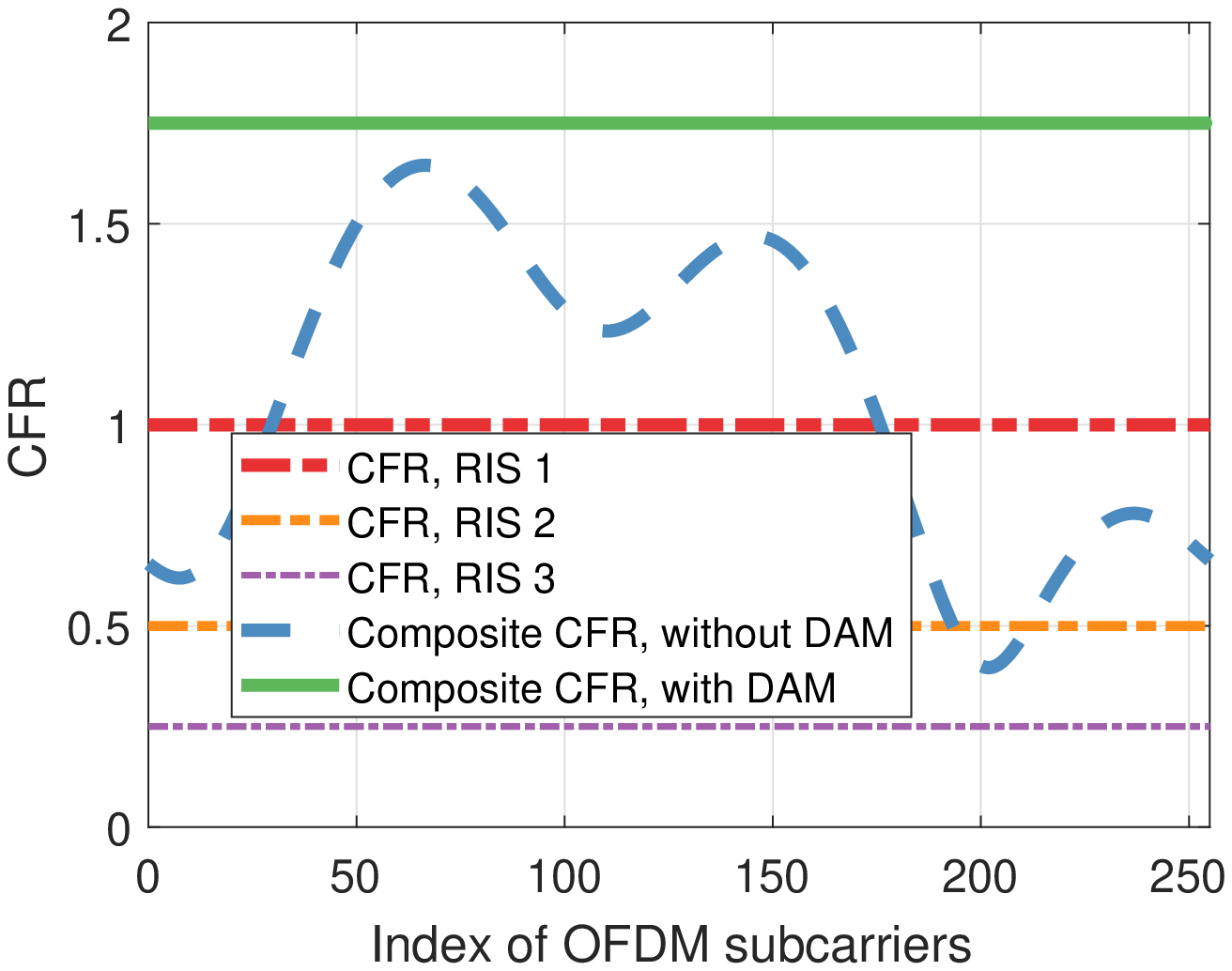}}
	\subfigure[\label{f3}]{\includegraphics[width=5.4cm]{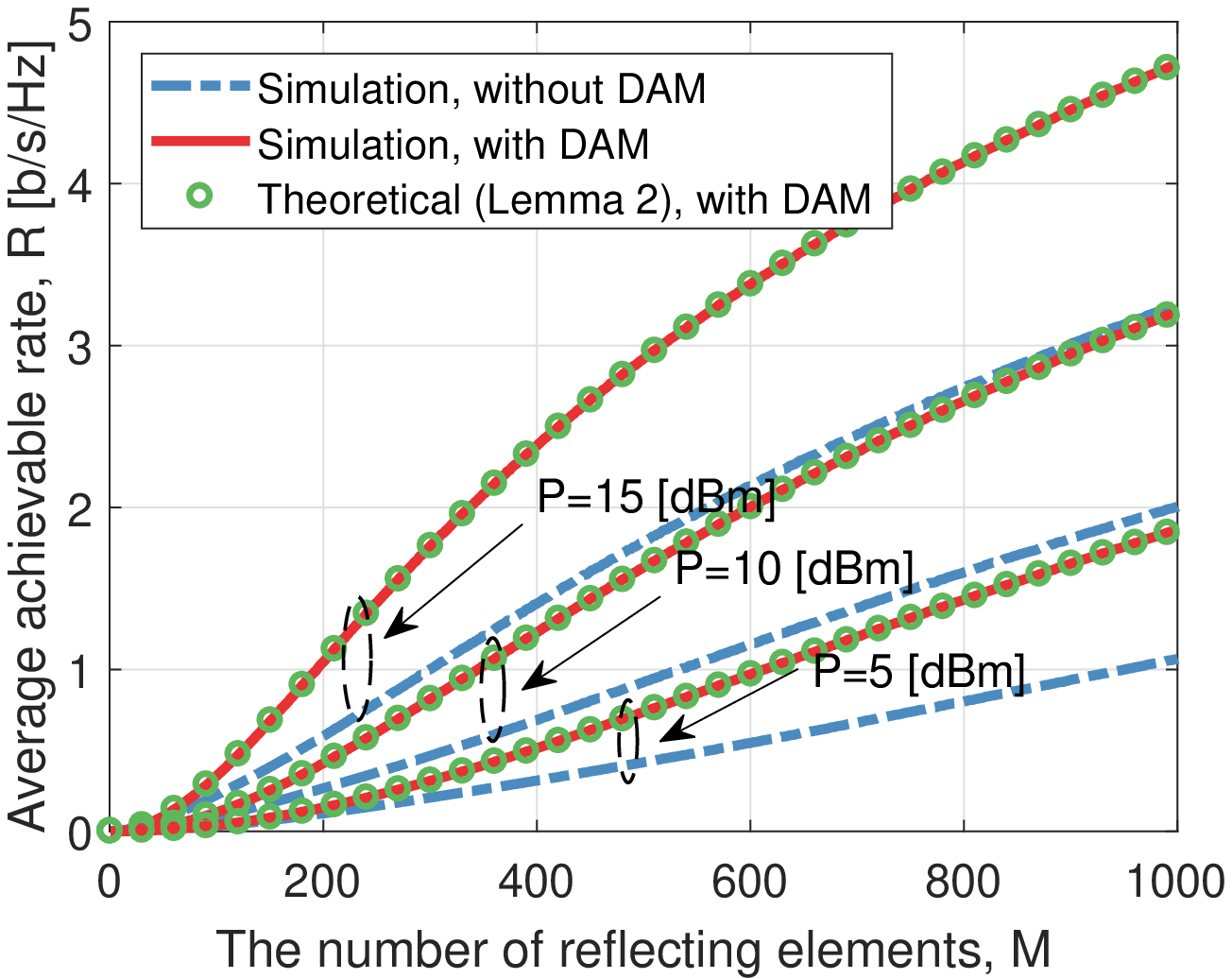}}
	\caption{(a) The comparison of the achievable rate of (\ref{eq12}) and its tight upper bound of (\ref{eq17}), where we have $K = 1$, $N = 1024$, ${N_{CP}} = 16$, ${\sigma ^2} =  - 100\ \text{dBm}$, $\Gamma  = 1$, ${\rho ^2} =  - 100\ \text{dB}$, $L = 3$. (b) The CFRs of three reflected channels and their composite channel. DAM enables the coherent superposition of three channels on all subcarriers. (c) The rate comparison of RIS-enhanced OFDM systems in the presence/absence of DAM, where we have $N = 1024$, ${N_{CP}} = 16$, ${\sigma ^2} =  - 80\ \text{dBm}$, $\Gamma  = 1$, $\rho _1^2 =  - 115\ \text{dB}$, ${L_1} = 1$, $\rho _2^2 =  - 120\ \text{dB}$, ${L_2} = 2$, $\rho _3^2 =  - 125\ \text{dB}$, ${L_3} = 4$.}
	\vspace{-0.8cm}
\end{figure}

For the sake of illustration, Fig. \ref{f2} compares the composite CFR in the presence and absence of DAM, where we consider three RISs, each equipped with a single reflecting element. Specifically, the reflected channels via three RISs are characterized by $\left| {{h_1}} \right| = 1$, $\angle {h_1} = 0$, ${L _1} = 1$; $\left| {{h_2}} \right| = \frac{1}{2}$, $\angle {h_2} = \frac{3}{4}\pi $, ${L _2} = 2$; $\left| {{h_3}} \right| = \frac{1}{4}$, $\angle {h_3} = \frac{3}{2}\pi$, ${L _3} = 4$. The number of OFDM subcarriers is set to $N = 256$. It can be shown from Fig. \ref{f2} that DAM is capable of facilitating coherent superposition of CFRs of three reflected channels, which provides a better channel quality and thus corresponds to a higher achievable rate. More specifically, the ergodic achievable rate of the RIS-enhanced OFDM systems employing DAM is summarized in \textbf{\emph{Lemma 2}}.\par
\textbf{\emph{Lemma 2:}} Assume all one-tap reflected channels satisfying ${h_{k,m}} \sim \mathcal{CN}\left( {0,{\rho _k^2}} \right),\ \forall m \in \mathcal{M},\ \forall k \in \mathcal{K}$. As $M \to \infty $, the ergodic achievable rate of the RIS-enhanced OFDM systems relying on DAM is given by\begin{small}\begin{align}
{\mathbb{E}}\left( R \right) = {\frac{N}{{N + {N_{CP}}}}}{\log _2}\left( {1 + \frac{{\left( {\pi {M^2}{{\left( {\sum\limits_{k = 1}^K {{\rho _k}} } \right)}^2} + \left( {4 - \pi } \right)M\sum\limits_{k = 1}^K {\rho _k^2} } \right)P}}{{4N\Gamma {\sigma ^2}}}} \right).
\end{align}\end{small}\par
\textbf{\emph{Proof:}} It can be readily shown that the composite channel remains the one-tap property upon leveraging the reflection coefficients and delays given in \textbf{\emph{Lemma 1}}, thus the composite CFR at each subcarrier can be denoted by
\begin{small}
\begin{align}
    {d_0} = {d_1} =  \cdots  = {d_{N-1}} = d = \sum\limits_{k = 1}^K {\sum\limits_{m = 1}^M {\left| {{h_{k,m}}} \right|} }.
\end{align}
\end{small}Hence, the WF algorithm reduces to the equal power allocation, i.e., ${p_0} =  \cdots  = {p_{N - 1}} = {P \mathord{\left/
 {\vphantom {P N}} \right.
 \kern-\nulldelimiterspace} N}$.\par
Moreover, as $M \to \infty$, we have\begin{small}\begin{align}
    d \sim {\mathcal{N}}\left( {\frac{{\sqrt \pi  }}{2}M\sum\limits_{k = 1}^K {{\rho _k}} ,\frac{{4 - \pi }}{4}M\sum\limits_{k = 1}^K {\rho _k^2} } \right),
\end{align}\end{small}according to the central limit theorem \cite{Montgomery_book_2011_Applied}. Therefore, the ergodic achievable rate of the RIS-enhanced OFDM systems employing DAM is given by\begin{small}\begin{align}
\mathbb{E}\left( R \right) &= \mathbb{E}\left( {\frac{N}{{N + {N_{CP}}}}{{\log }_2}\left( {1 + \frac{{{d^2}P}}{{N\Gamma {\sigma ^2}}}} \right)} \right) \stackrel{M \to \infty}{\longrightarrow}\frac{N}{{N + {N_{CP}}}}{\log _2}\left( {1 + \frac{{\mathbb{E}\left( {{d^2}} \right)P}}{{N\Gamma {\sigma ^2}}}} \right) \notag\\
 &= \frac{N}{{N + {N_{CP}}}}{\log _2}\left( {1 + \frac{{\left( {\pi {M^2}{{\left( {\sum\limits_{k = 1}^K {{\rho _k}} } \right)}^2} + \left( {4 - \pi } \right)M\sum\limits_{k = 1}^K {\rho _k^2} } \right)P}}{{4N\Gamma {\sigma ^2}}}} \right).
\end{align}\end{small}The proof is completed. $\hfill\blacksquare$\par
\textbf{\emph{Lemma 2}} demonstrates that the RIS-enhanced OFDM systems employing DAM remain the quadratic law with $M$ as for narrowband signals \cite{Wu_TWC_2019_Intelligent, wu_TC_2020_Beamforming}. For the sake of elaboration, Fig. \ref{f3} verifies the accuracy of \textbf{\emph{Lemma 2}} and compares the ergodic achievable rate of RIS-assisted OFDM systems in the presence and absence of DAM, where the specific simulation parameters are listed in Fig. \ref{f3}. It can be observed that the ergodic achievable rate of RIS-enhanced OFDM employing DAM outperforms that relying on the traditional RIS for nearly 50\%. Additionally, the theoretical analysis of \textbf{\emph{Lemma 2}} nicely matches the ergodic achievable rate of RIS-assisted OFDM systems relying on DAM, even for the scenarios that RISs involved are equipped with a small number of reflecting elements.
\section{Simulation Results}\label{S6}
\begin{table*}[!t]
\scriptsize
\centering
\caption{The operation setup of 12 benchmark schemes.}
\label{tab5}
\begin{tabular}{c|l|l|l|l||c|l|l|l|l}
\hline \hline
Scheme & Power allocation & DAM & Phase shift  & Reference & Scheme& Power allocation & DAM & Phase shift  & Reference   \\ \hline
1   & Equal PA solution & NO  & Random      & \diagbox[width=5.5em]{}{}     & 7    & WF solution & NO  & Random      & \diagbox[width=6.8em]{}{}            \\ \hline
2   & Equal PA solution & NO  & Statistical & \diagbox[width=5.5em]{}{}     & 8    & WF solution & NO  & Statistical &  \cellcolor{mygray}\cite{Han_TVT_2019_Large}           \\ \hline
3   & Equal PA solution & NO  & Optimal     & \cellcolor{mygray}\cite{Zheng_WCL_2020_Intelligent}     & 9    & WF solution & NO  & Optimal     & \cellcolor{mygray}\cite{Yang_TC_2020_Intelligent}            \\ \hline
4   & Equal PA solution & YES & Random      & \diagbox[width=5.5em]{}{}     & 10   & WF solution & YES & Random      &  \diagbox[width=6.8em]{}{}            \\ \hline
5   & Equal PA solution & YES & Statistical & \diagbox[width=5.5em]{}{}     & 11   & WF solution & YES & Statistical &   \diagbox[width=6.8em]{}{}           \\ \hline
6   & Equal PA solution & YES & Optimal     & \diagbox[width=5.5em]{}{}     & 12   & WF solution & YES & Optimal     & \cellcolor{mygray}\textbf{Algorithm 2} \\ \hline \hline
\end{tabular}
	\vspace{-0.8cm}
\end{table*}
\subsection{Simulation Setup}
\begin{figure}[!t]
	\centering
	\includegraphics[width=7cm]{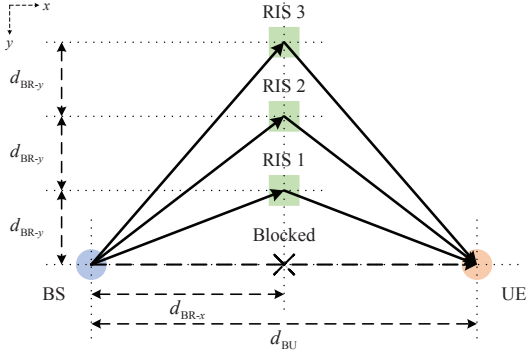}
	\caption{The position setting of the BS, UE, and RISs (from the top view).}
	\label{f4}
		\vspace{-0.8cm}
\end{figure}

In this section, we evaluate the performance of our proposed design and algorithms via numerical simulations. As shown in Fig. \ref{f4}, we consider a downlink OFDM system assisted by three RISs, all of which are assumed to be an uniform rectangular array consisted of $M$ reflecting elements with half-wavelength spacing. The direct BS-UE link is assumed to be blocked. For ease of exposition, we assume all RISs are placed along the $x$-$z$ plane and perpendicular to the ground, i.e., the $x$-$y$ plane. The reference element's location of the $k$-th RIS is set to $\left( {{d_{{\rm{BR}} - x}},-k{d_{{\rm{BR}} - y}},{d_{{\rm{R}} - z}}} \right),\ \forall k \in \mathcal{K}$, while the antennas' locations of the BS and the UE are set to $\left( {0,0,{d_{{\rm{B}} - z}}} \right)$ and $\left( {{d_{{\rm{BU}} - x}},0,0} \right)$, respectively. In our simulations, the horizontal distance between the BS and all RISs is set to ${d_{{\rm{BR}}-x}} = 100\ \text{m}$. The distance between different RISs is set to ${d_{{\rm{BR}} - y}} = 20\ \text{m}$, while the height of the BS and all RISs is set to ${d_{{\rm{B}} - z}} = {d_{{\rm{R}} - z}} = 10\ \text{m}$. Note that the locations of the practical RIS deployment are generally chosen to favor LoS propagation between the RISs and the BS/UE. Therefore, for each RIS, the separate BS-RIS and RIS-UE links are both modeled by multi-path Rician fading channels with the first non-zero tap of each channel being the deterministic LoS path and the remaining non-zero taps characterizing the non-LoS (NLoS) paths \cite{Foerster_TC_1997_Analysis}. Explicitly, the maximum delay spread of the $k$-th BS-RIS link is set to ${L_{k,{\rm{BR}}}} = L_{k,{\rm{BR}}}^{{\rm{zero}}} + L_{k,{\rm{BR}}}^{{\rm{non - zero}}},\ \forall k \in \mathcal{K}$ taps, where the first $L_{k,{\rm{BR}}}^{{\rm{zero}}} = \left\lfloor {{{{d_{k,{\rm{BR}}}}{f_s}} \mathord{\left/
 {\vphantom {{{d_{k,{\rm{BR}}}}{f_s}} c}} \right.
 \kern-\nulldelimiterspace} c}} \right\rceil$ taps characterize the transmission delay of the BS-RIS link associated with the $k$-th RIS, while the last $L_{k,{\rm{BR}}}^{{\rm{non - zero}}}$ taps depend on the specific scatter environment of the $k$-th BS-RIS channel. For the sake of illustration, we set $L_{k,{\rm{BR}}}^{{\rm{non - zero}}}=1$ with Rician factor of ${\zeta _{k,{\rm{BR}}}} = \infty,\ \forall k \in \mathcal{K} $ (e.g., a macro BS and multiple RISs attached to the skyscrapers' upper-layer surface). The sampling rate is set to ${f_s} = 50\ \text{MHz}$. Similarly, the maximum delay spread of the $k$-th RIS-UE link is set to ${L_{k,{\rm{RU}}}} = L_{k,{\rm{RU}}}^{{\rm{zero}}} + L_{k,{\rm{RU}}}^{{\rm{non - zero}}},\ \forall k \in \mathcal{K}$ taps, where the first $L_{k,{\rm{RU}}}^{{\rm{zero}}} = \left\lfloor {{{{d_{k,{\rm{RU}}}}{f_s}} \mathord{\left/
 {\vphantom {{{d_{k,{\rm{RU}}}}{f_s}} c}} \right.
 \kern-\nulldelimiterspace} c}} \right\rceil $ taps characterize the transmission delay of the RIS-UE link associated with the $k$-th RIS, while the last $L_{k,{\rm{RU}}}^{{\rm{non - zero}}}$ taps characterize the $k$-th RIS-UE channel depending on the surrounding scatters. Without other specified, we set $L_{k,{\rm{RU}}}^{{\rm{non - zero}}} =5$ taps, each non-zero NLoS tap coefficient is modeled by Rayleigh fading with an uniform power delay profile, i.e., ${v_{k,m}}\left( l \right) \sim \mathcal{CN}\left( {0,1} \right),\;l = L_{k,{\rm{RU}}}^{{\rm{zero}}},L_{k,{\rm{RU}}}^{{\rm{zero}}} + 1, \cdots ,{L_{k,{\rm{RU}}}} - 1$ \cite{Foerster_TC_1997_Analysis}. The Rician factor of all RIS-UE links is set to ${\zeta _{k,{\rm{RU}}}} = 3\ {\rm{dB}},\ \forall k \in \mathcal{K}$. In order to maintain the channel gain normalization, the generated RIS-UE channels are then scaled by multiplying a coefficient of ${1 \mathord{\left/
 {\vphantom {1 {\sqrt {{\zeta _{k,{\rm{RU}}}} + L_{k,{\rm{RU}}}^{{\rm{non - zero}}}} }}} \right.
 \kern-\nulldelimiterspace} {\sqrt {{\zeta _{k,{\rm{RU}}}} + L_{k,{\rm{RU}}}^{{\rm{non - zero}}}} }}$.\par
Furthermore, for the LoS path, as the RIS array size is practically much smaller than the distance of the BS-RIS and RIS-UE links, the channel gains of all links reflected via the same RIS are approximately identical while their phases are correlated depending on the respective physical layout. The rays arriving at the $k$-th RIS are assumed to be parallel for all elements, with a common angle of arrival (AOA) composed of an elevation angle of ${\varphi _{e,k,{\rm{BR}}}} = 0$ and an azimuth angle of ${\varphi _{a,k,{\rm{BR}}}} = \arctan \left( {{{k{d_{{\rm{BR}} - y}}} \mathord{\left/
 {\vphantom {{k{d_{{\rm{BR}} - y}}} {{d_{{\rm{BR}} - x}}}}} \right.
 \kern-\nulldelimiterspace} {{d_{{\rm{BR}} - x}}}}} \right),\ \forall k \in \mathcal{K}$. Similarly, the reflected rays leaving the $k$-th RIS follow a common angle of departure (AOD) composed of an elevation angle of ${\varphi _{e,k,{\rm{RU}}}} =  - \arctan \left( {{{{d_{{{\rm{R}}} - z}}} \mathord{\left/
 {\vphantom {{{d_{{{\rm{R}}} - z}}} {\sqrt {{{\left( {{d_{{\rm{BU}} - x}} - {d_{{\rm{BR}} - x}}} \right)}^2} + {{\left( {k{d_{{\rm{BR}} - y}}} \right)}^2}} }}} \right.
 \kern-\nulldelimiterspace} {\sqrt {{{\left( {{d_{{\rm{BU}} - x}} - {d_{{\rm{BR}} - x}}} \right)}^2} + {{\left( {k{d_{{\rm{BR}} - y}}} \right)}^2}} }}} \right)$ and an azimuth angle of ${\varphi _{a,k,{\rm{RU}}}} = {\pi  \mathord{\left/
 {\vphantom {\pi  2}} \right.
 \kern-\nulldelimiterspace} 2}{\rm{ + }}\arctan \left( {{{\left( {{d_{{\rm{BU}} - x}} - {d_{{\rm{BR}} - x}}} \right)} \mathord{\left/
 {\vphantom {{\left( {{d_{{\rm{BU}} - x}} - {d_{{\rm{BR}} - x}}} \right)} {k{d_{{\rm{BR}} - y}}}}} \right.
 \kern-\nulldelimiterspace} {k{d_{{\rm{BR}} - y}}}}} \right)$. For an arbitrary one of RISs, let $\left( {{m_x},{m_z}} \right)$ denote the location of a RIS element, with $1 \le {m_x} \le {M_x}$ and $1 \le {m_z} \le {M_z}$. In the following simulations with varying $M$, we fix ${M_x} = 10$ and increase $M_z$ linearly with $M = {M_x}{M_z}$. Let $\omega \left( {{m_x},{m_z}} \right)$ denote the phase offset of the BS-RIS link at $\left( {{m_x},{m_z}} \right)$ with respect to that at $\left( {1,1} \right)$ (i.e., the one at the top left), thus we have\begin{small}\begin{align}
{\omega _{k,{\rm{BR}}}}\left( {{m_x},{m_z}} \right) = \frac{{2\pi }}{\lambda }\left( {{m_x} - 1} \right)d\cos {\varphi _{e,k,{\rm{BR}}}}\cos {\varphi _{a,k,{\rm{BR}}}} + \frac{{2\pi }}{\lambda }\left( {{m_z} - 1} \right)d\sin {\varphi _{e,k,{\rm{BR}}}}, \label{eq35}
\end{align}\end{small}where $d$ denotes the RIS element spacing and $\lambda $ denotes the carrier wavelength. In our simulations, we set $d = 0.05\ \text{m}$ and $\lambda  = 0.33\ \text{m}$ corresponding to a carrier frequency of ${f_c} = 900\ \text{MHz}$ \cite{Tang_TWC_2021_Wireless}. Similarly, the phase offset for the LoS path of the RIS-UE link ${\omega _{k,{\rm{RU}}}}\left( {{m_x},{m_z}} \right)$ can be obtained by substituting ${\varphi _{e,k,{\rm{RU}}}}$ and ${\varphi _{a,k,{\rm{RU}}}}$ into (\ref{eq35}). As a result, the phase difference of the LoS paths for all the RIS elements in the BS-RIS link and RIS-UE link are fixed. Moreover, the path loss of the BS-RIS and RIS-UE links is modeled by $\rho _{k,{\rm{BR}}/{\rm{RU}}}^2 = {C_0}d_{k,{\rm{BR}}/{\rm{RU}}}^{ - {\alpha _{k,{\rm{BR}}/{\rm{RU}}}}}$, where ${C_0}=-30$ dB is the power loss at the reference distance of $d = 1\ \text{m}$, ${d_{k,\rm{BR}}}$ and ${d_{k,\rm{RU}}}$ denote the distance of the BS-RIS and the RIS-UE links, respectively, with respect to the $k$-th RIS; while ${\alpha _{k,\rm{BR}}}=2.2$ and ${\alpha _{k,\rm{RU}}}=2.8$ denote the corresponding path loss exponents, respectively, which are assumed to be the same for all RISs. Additionally, the number of OFDM subcarriers is set to $N=1024$, while the CP length is thus set as ${N_{CP}} = 16$. In our simulations, we assume ${\bar \tau _{\max }} \ge {N_{CP}}$ in order to excavate the best performance of DAM. The total transmit power at the BS is given by $P = 20\ \text{dBm}$, while the average noise power is set to ${\sigma ^2} =  - 100\ \text{dBm}$ \cite{an_arxiv_2021_joint}. In order to characterize the maximum achievable rate, the SNR gap is set as $\Gamma  = 0$ dB. The number of Gaussian randomization for solving Problem $\left( {P4\text{-SDR}} \right)$ in \textbf{Algorithm 1} is chosen as $Q = 100$, while the number of random initializations in \textbf{Algorithm 1} is set to $J = 10$. All the results are averaged over $1000$ independent channel realizations.\par
\subsection{Benchmark Schemes}
Firstly, we assume that perfect CSI is available and ignore any pilot overhead for CSI acquisition. The achievable rate is thus computed using (\ref{eq12}). For the sake of comparison, we consider the following $12$ benchmark schemes upon considering two transmit power allocation schemes (i.e., equal solution and WF solution), two kinds of RIS elements (i.e., employing DAM or not) and three phase shift configuration methods (i.e., random, statistical, optimal). More specifically, the $12$ benchmark schemes are listed in Table \ref{tab5}, where their respective operation setups are demonstrated as follows:\par
\emph{1) Transmit power allocation}
\begin{itemize}
\item Equal PA solution: Allocating an equal amount of power to all subcarriers;
\item WF solution: Performing the power allocation according to the WF solution of (\ref{eq15}).
\end{itemize}\par
\emph{2) RIS elements}
\begin{itemize}
\item Employing no DAM: Each reflecting element is capable of only adjusting the phase of the incident signals;
\item Employing DAM: Each reflecting element is capable of not only adjusting the phase of the incident signals but imposing an extra delay.
\end{itemize}\par
\emph{3) Phase shift configuration method}
\begin{itemize}
\item Random: Each RIS coefficient has a random phase independently and uniformly distributed in $\left[ {0,2\pi } \right)$;
\item Statistical: Each RIS coefficient are configured by aligning the determined LOS components of the BS-RIS and RIS-UE links;
\item Optimal\footnote{Note that here the optimal method refers to the optimal phase shift configuration required to align all taps having the maximum gain, not the optimal RIS configuration in the global sense. The optimal design described hereinafter follows the same consideration.}: Each RIS coefficient are optimally configured based on (\ref{eq27}) in \textbf{Algorithm 2}.
\end{itemize}Note that for the RIS elements employing DAM, the statistical phase shift configuration method adjusts the RIS's delay aligning to the LoS path of each reflected channel, while the optimal phase shift configuration method adjusts the RIS's delay according to (\ref{eq27}).\par
\subsection{Performance Comparison with Benchmark Schemes}
\begin{figure}[!t]
	\centering
	\subfigure[\label{f5}]{\includegraphics[width=5.4cm]{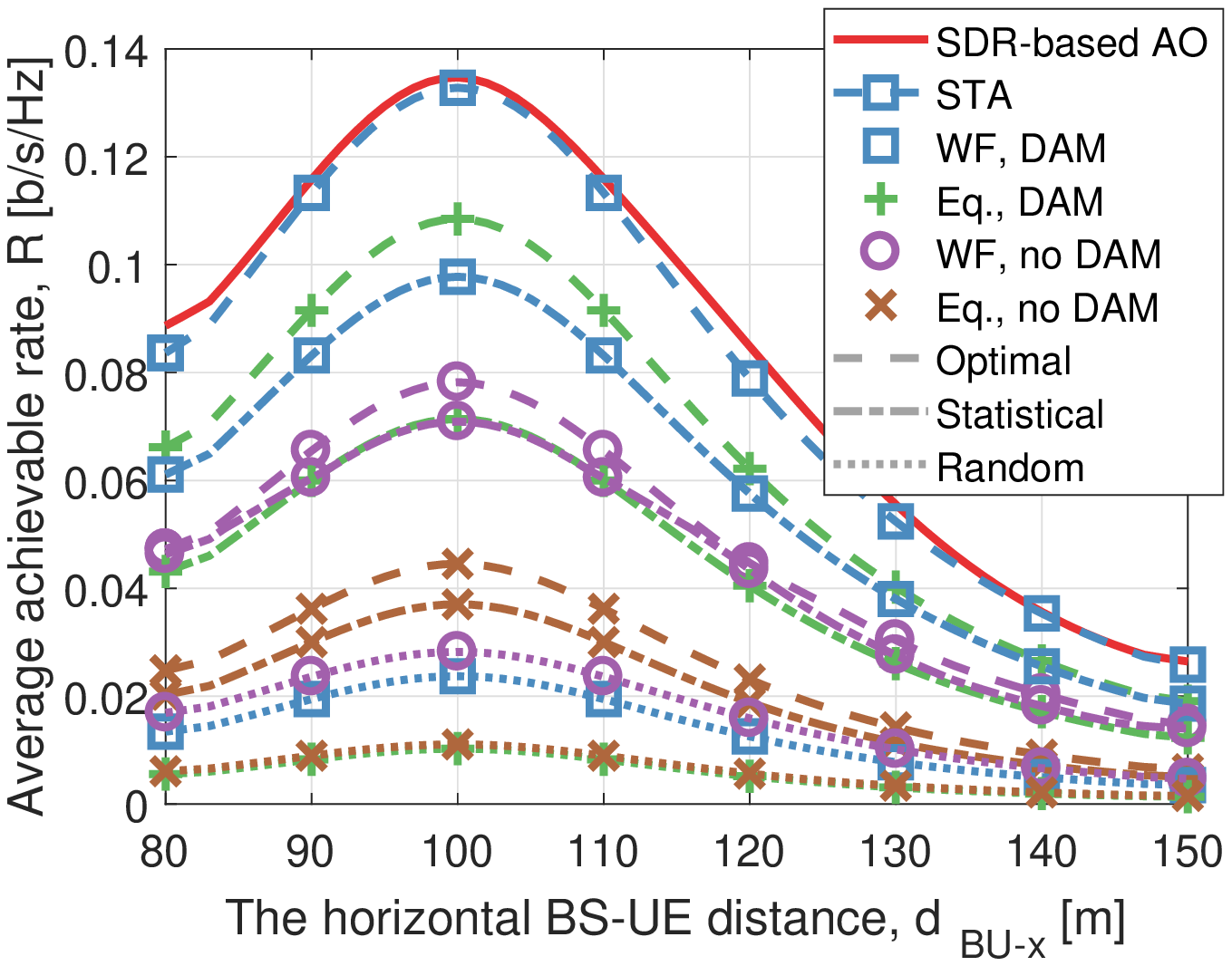}}
	\subfigure[\label{f6}]{\includegraphics[width=5.4cm]{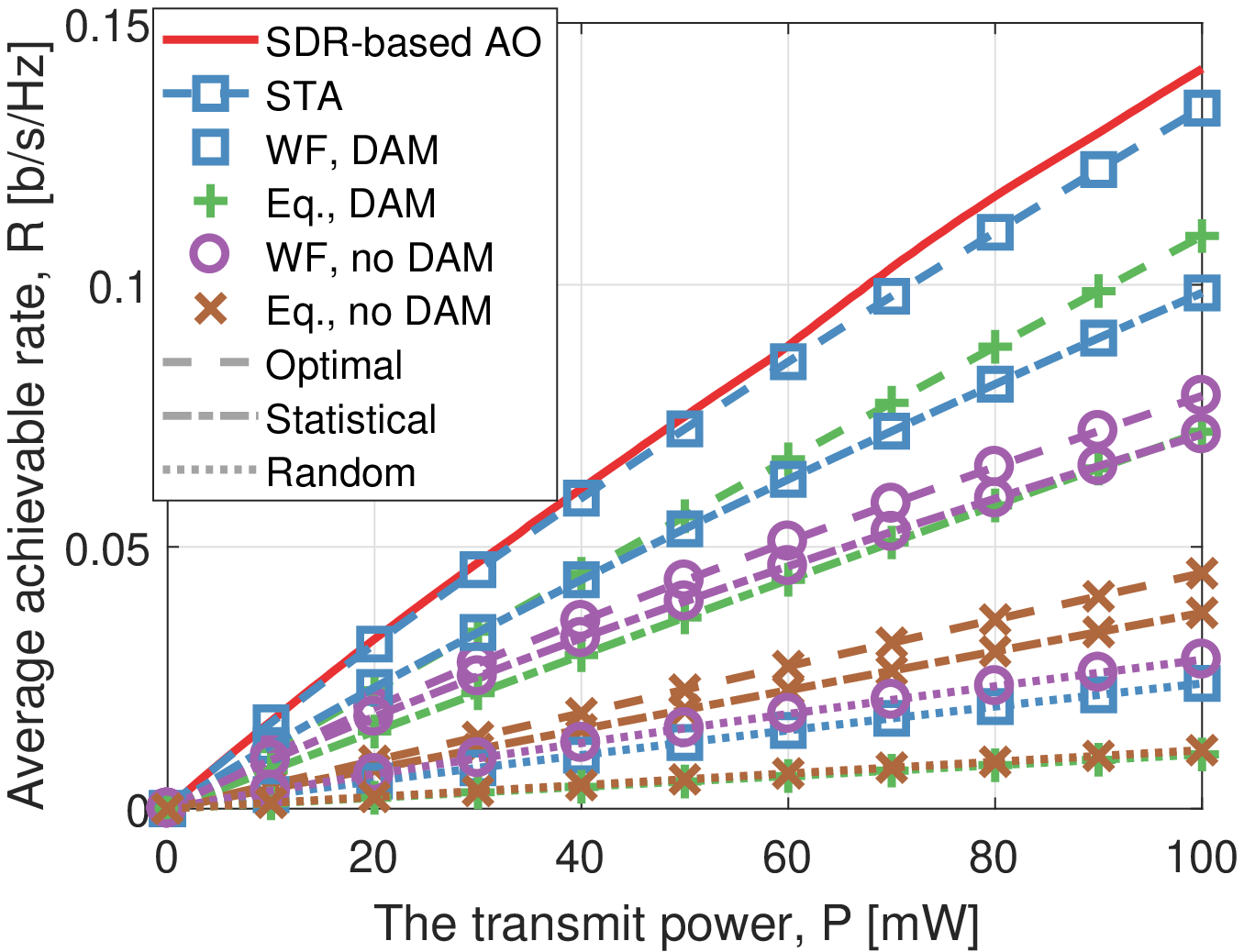}}
	\caption{(a) Achievable rate versus the horizontal BS-UE distance. (b) Achievable rate versus the total transmit power.}
	\vspace{-0.8cm}
\end{figure}
\begin{figure}[!t]
	\centering
	\subfigure[\label{f7}]{\includegraphics[width=5.4cm]{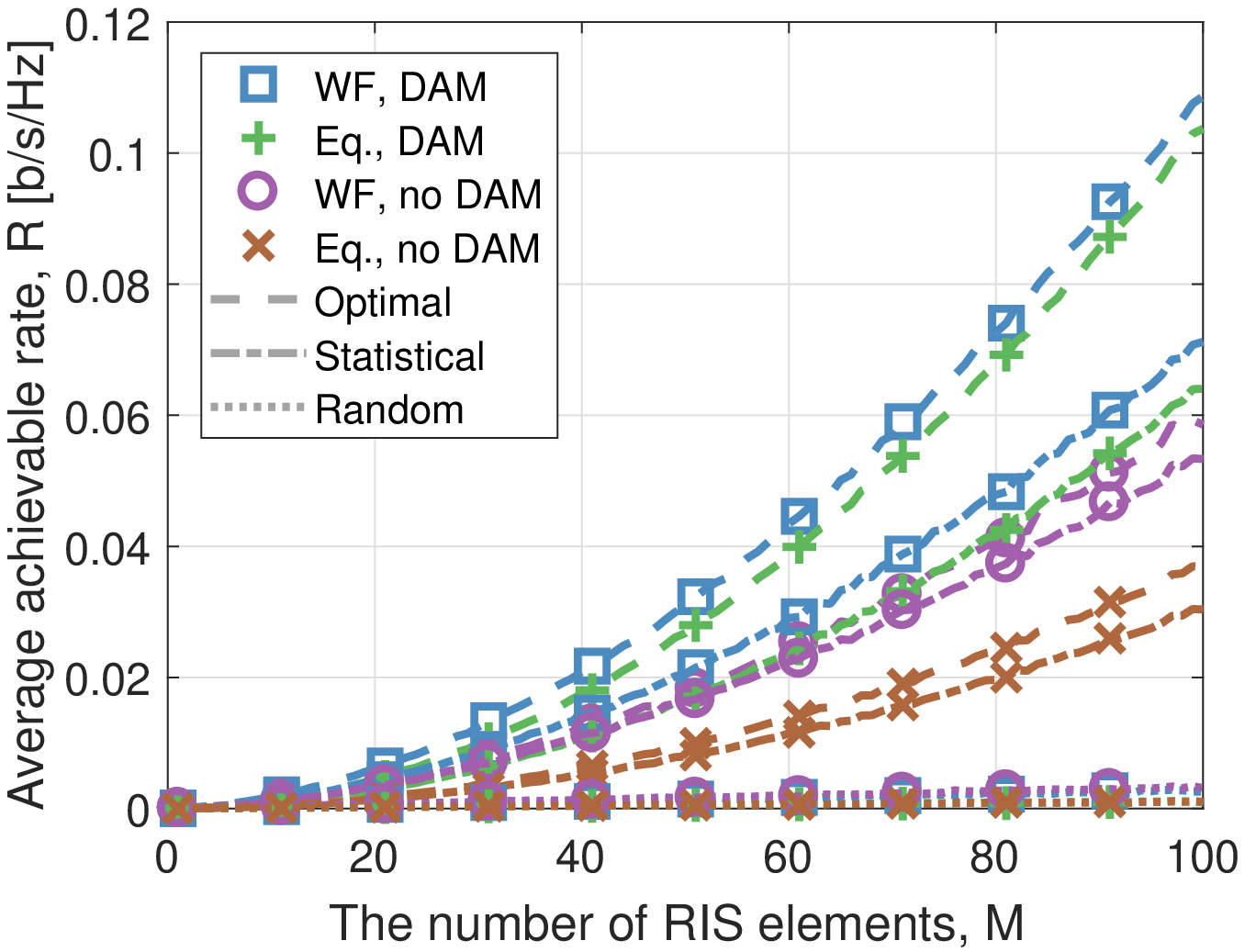}}
	\subfigure[\label{f8}]{\includegraphics[width=5.4cm]{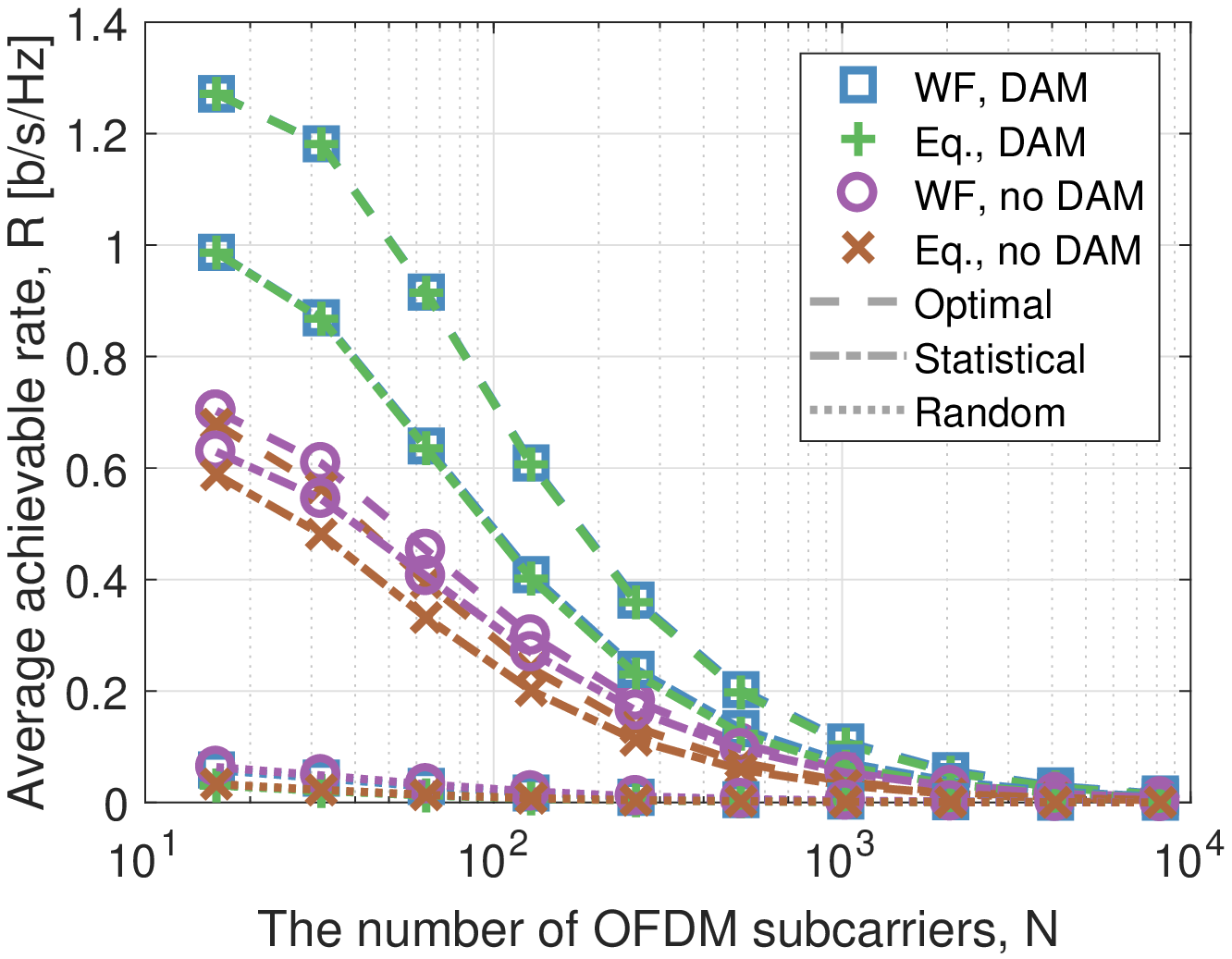}}
	\caption{(a) Achievable rate versus the number of RIS elements. (b) Achievable rate versus the number of OFDM subcarriers.}
	\vspace{-0.8cm}
\end{figure}
Fig. \ref{f5} depicts the achievable rate performance versus the horizontal BS-UE distance, where we consider ${M_z} = 1$. It can be observed from Fig. \ref{f5} that for random-configured RIS phase shifts, the employment of DAM hardly brings any performance improvement. Furthermore, for the conventional RIS-assisted OFDM systems, the phase shift configuration based on the statistical CSI (i.e., the deterministic LoS path) and that based on the instantaneous CSI result in improved rate performance, following which the optimal WF power allocation solution can further increase the achievable rate compared to their corresponding counterparts relying on equal power allocation solution. Moreover, DAM is capable of substantially lifting the achievable rate over those without employing DAM upon aligning the strongest taps of different reflected channels by adjusting the RIS delay. Explicitly, when the UE moves at the nearest vicinity of RIS, i.e., ${d_{{\rm{BU}} - x}} = 100\ \text{m}$, the employment of DAM increases the achievable rate from $0.08\ \text{b/s/Hz}$ to $0.135\ \text{b/s/Hz}$ for the optimal phase shift configuration relying on WF solution, with a rate increase of about $70\%$. In summary, RIS is capable of creating a hot spot by collecting energy diffused in the channel to improve the QoS of users in its vicinity, and the DAM introduced in this paper can further stimulate the potential of RIS for wideband communications. Besides, the conceived STA method almost matches the high-quality sub-optimal \textbf{Algorithm 1}, albeit its significantly reduced complexity.\par
In Fig. \ref{f6}, we compare the achievable rate of the SDR-based AO method and benchmark schemes versus the transmit power at the BS, where we set ${M_{z}} = 1$ and $d_{{\rm{BU}}-x} = 100\ \text{m}$. As expected, the achievable rate increases with the transmit power. More specifically, the SDR-based AO method and the conceived STA method preserve the fastest growth. The low-complexity STA method behaves competitively with respect to the SDR-based AO algorithm, with only a rate erosion of about $0.005\ \text{b/s/Hz}$ for a large amount of transmit power. Compared to the conventional RIS-assisted OFDM systems relying on the optimal configuration, the employment of DAM increases the achievable rate from $0.075\ \text{b/s/Hz}$ to $0.13\ \text{b/s/Hz}$ for $P = 100 \text{mW}$, maintaining a rate increase of more than $70\%$, which benefits from the fact that the introduced DAM performs beam-focusing on all subcarriers rather than on the carrier frequency as in the conventional design, which significantly improves the beamforming efficiency. In addition, for the conventional RIS-assisted OFDM systems, the statistical phase shift configuration almost matches the optimal phase shift design with only $0.01\ \text{b/s/Hz}$ rate loss. By contrast, the DAM exploits the instantaneous CSI more effectively by increasing the rate by about $0.04\ \text{b/s/Hz}$, whether for the power allocation relying on WF solution or the equal PA solution, which is also due to the fact that delay adjustable RIS further improves the composite channel gain. In a nutshell, the setups employing DAM outperforms their conventional counterparts.\par
Fig. \ref{f7} compares the rate performance of different benchmark schemes versus the number of RIS elements. Note that it has been shown that the proposed STA method incurs only marginal performance loss yet with much lower complexity compared to the SDR-based AO algorithm. Thus, we consider only the suboptimal STA method in the following in order to bypass the excessive computational burden of the SDR-based AO method for a large value of $M$. Observe from Fig. \ref{f7} that even the proposed DAM-oriented scheme adopting the equal power allocation solution outperforms all non-DAM-oriented counterparts. Specifically, for the conventional RIS-assisted systems, the optimal RIS configuration relying on WF solution barely catch up with the statistical RIS configuration employing DAM. The equal power allocation solution without DAM results in poorer performance. Furthermore, note that as the number of elements increases, the advantages of the proposed design over its non-DAM-oriented counterparts will be gradually highlighted, which is because the RIS employing DAM is capable of collecting more power on the aligned tap for an increased number of RIS elements.\par
\begin{figure}[!t]
	\centering
	\subfigure[\label{f9}]{\includegraphics[width=5.4cm]{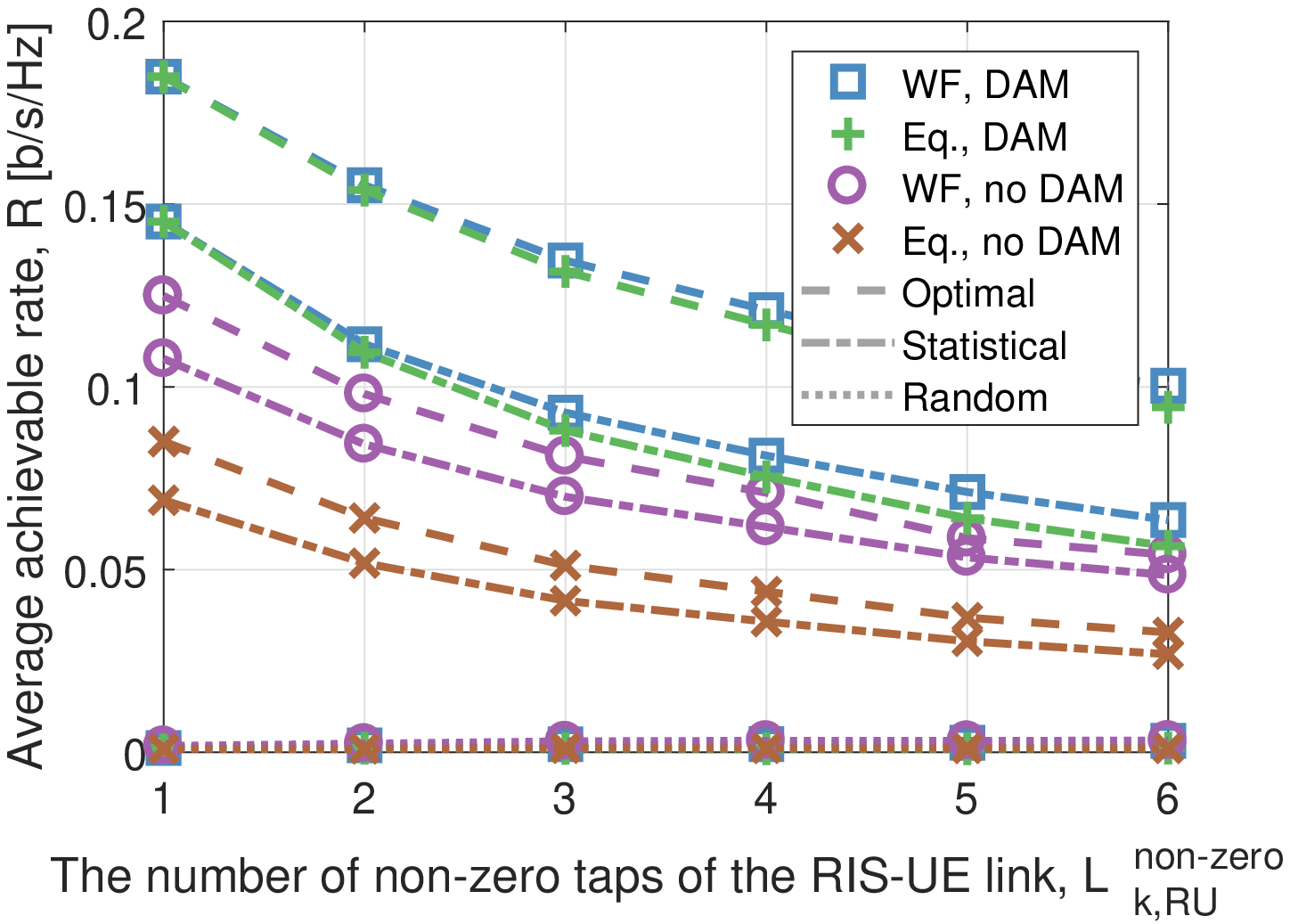}}
	\subfigure[\label{f10}]{\includegraphics[width=5.4cm]{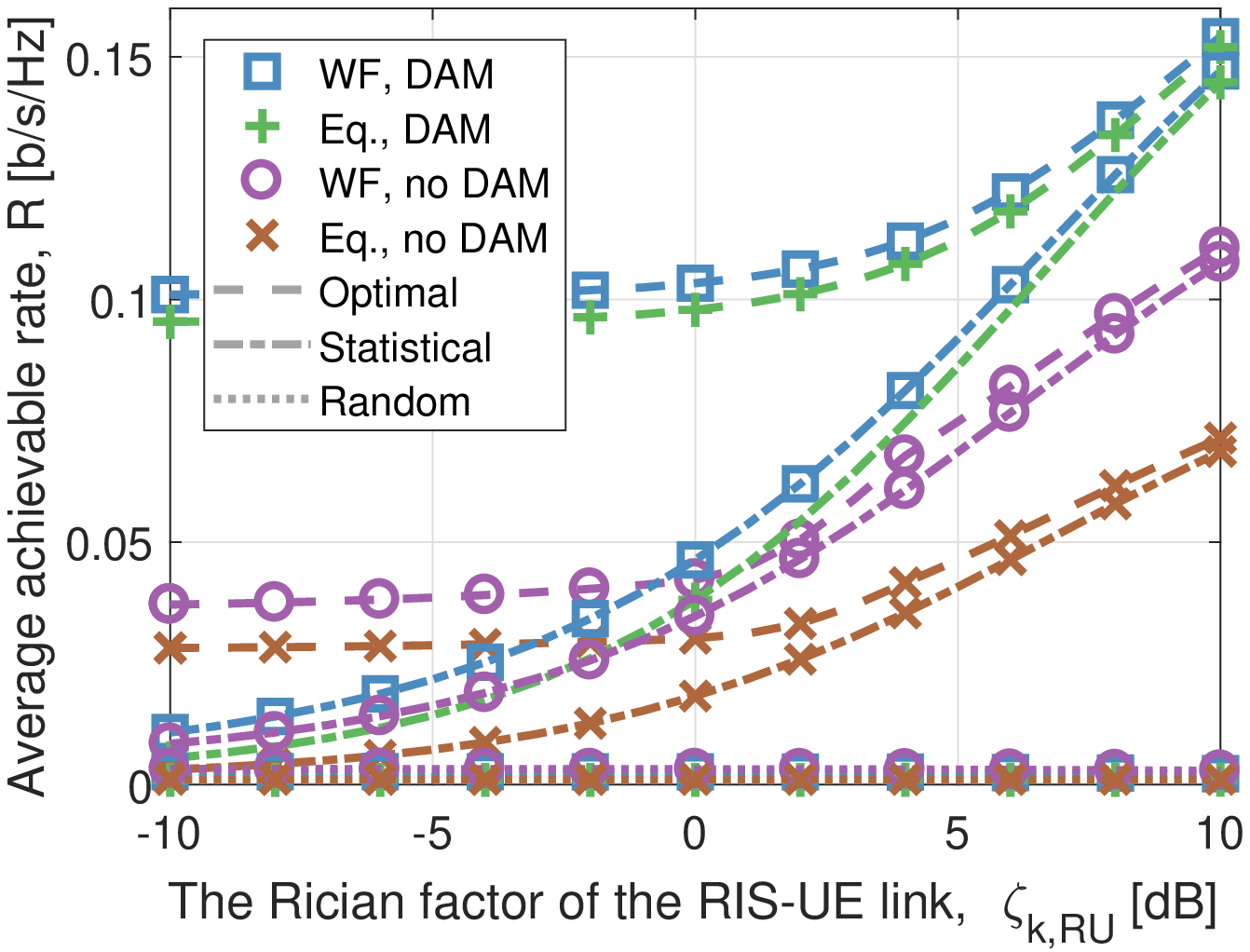}}
	\caption{(a) Achievable rate versus the number of non-zero taps of the RIS-UE link. (b) Achievable rate versus the Rician factor.}
		\vspace{-0.8cm}
\end{figure}
\begin{figure}[!t]
	\centering
	\subfigure[]{\includegraphics[width=5.4cm]{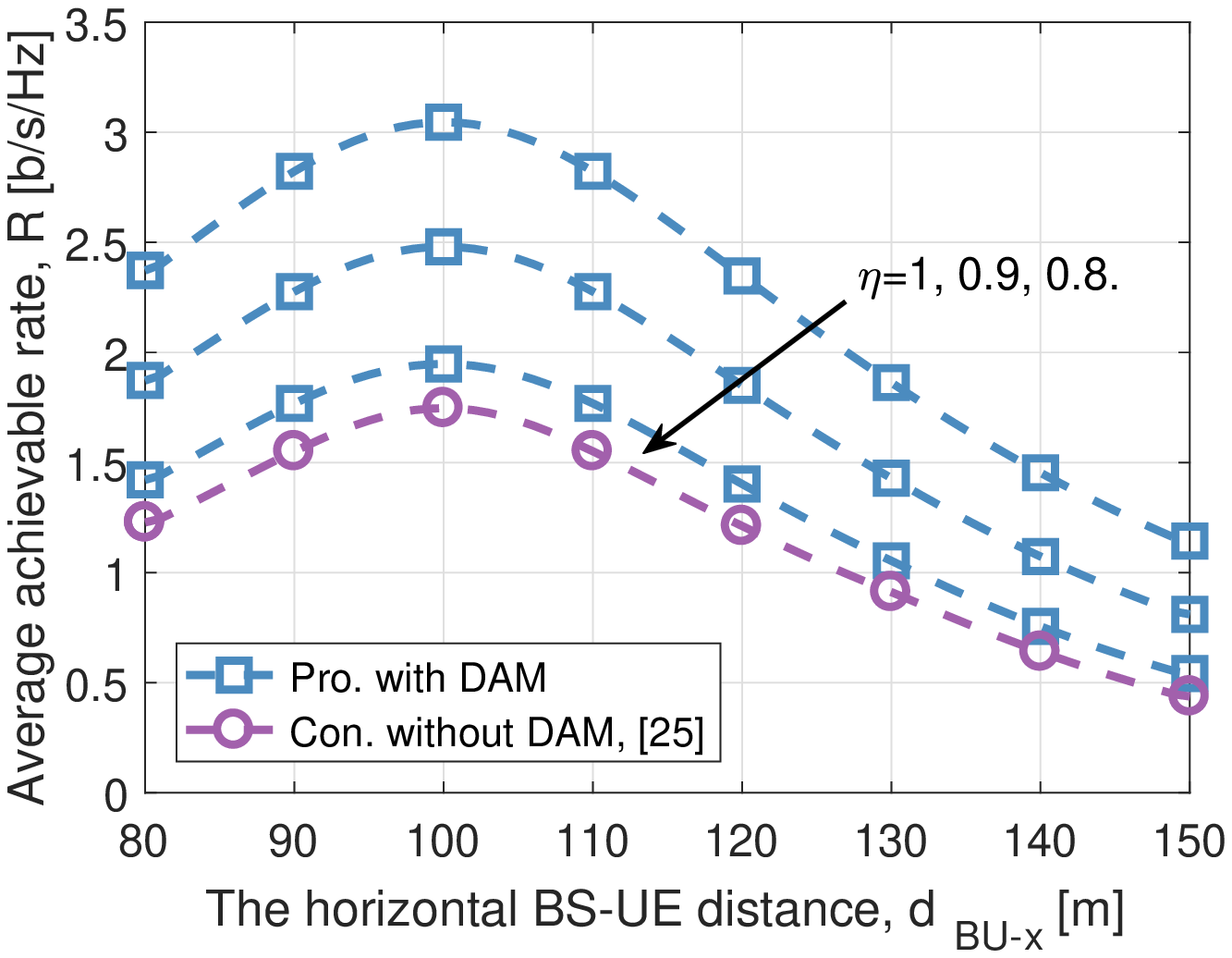}}
	\subfigure[]{\includegraphics[width=5.4cm]{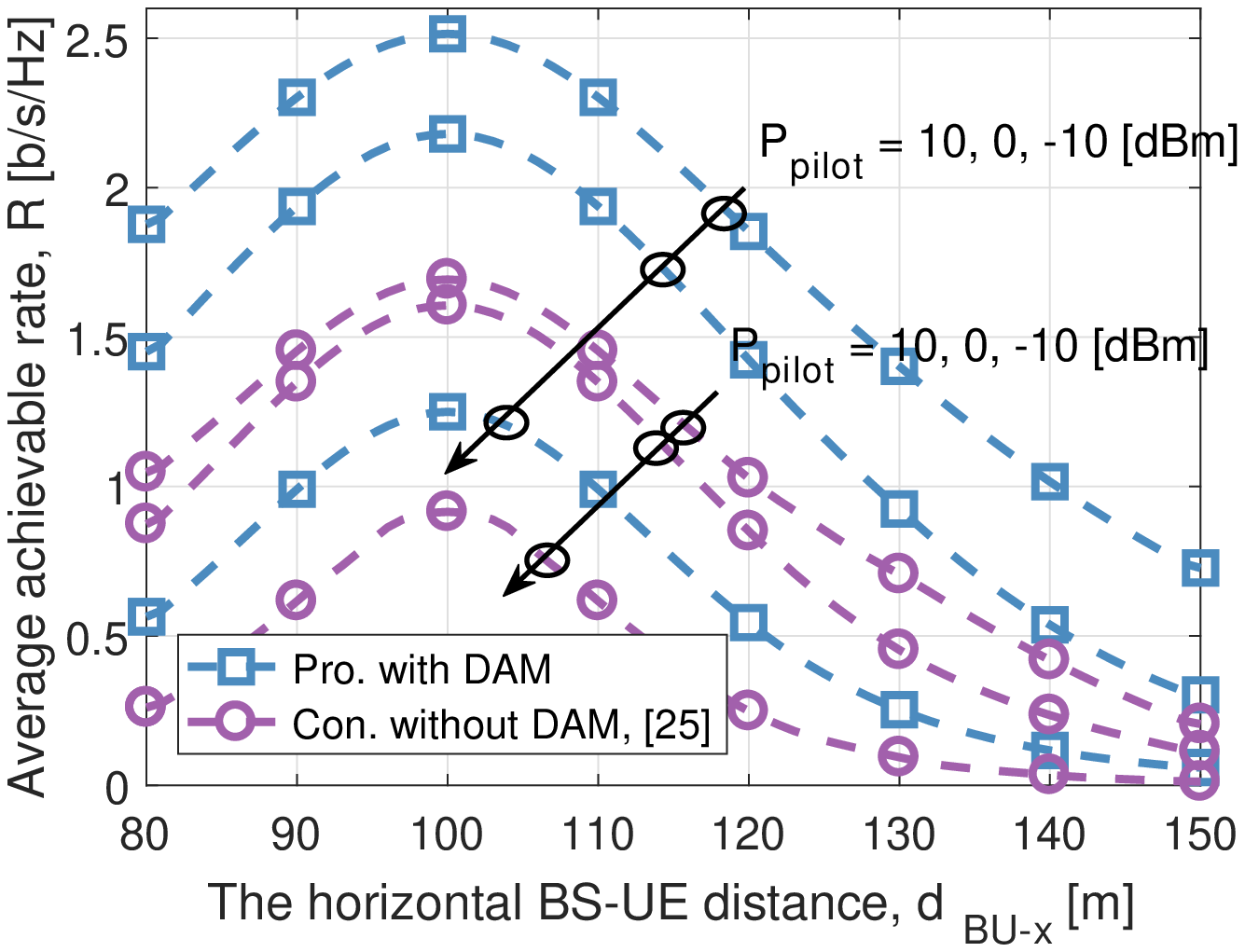}}
	\caption{(a) Achievable rate versus the horizontal BS-UE distance, where the practical DAM component's power decay is considered. (b) Achievable rate versus the horizontal BS-UE distance, where the channel estimation errors are considered. The decay factor is set to $\eta  = 0.9$.}
	\label{fig11}
	\vspace{-0.8cm}
\end{figure}
Fig. \ref{f8} portrays the achievable rate versus the number of OFDM subcarriers, where we set ${M_z}=10$. It can be seen from Fig. \ref{f8} that the achievable rate decreases with the growing number of subcarriers, which potentially implies that the RIS prefers to bring higher performance gain to the narrowband signals. Hence, it remains an open question to adopt the single-carrier or multi-carrier transmission scheme for future large-scale RIS deployment. Nevertheless, the employment of DAM always brings significant performance improvement, no matter adopting the optimal RIS configuration relying on instantaneous CSI or the suboptimal RIS configuration relying on statistical CSI. By contrast, the statistical RIS configuration without employing DAM performs competitively to its optimal RIS configuration based counterparts, both of which, however, suffer from a rate loss of about $0.55\ \text{b/s/Hz}$ than the proposed sub-optimal design when considering the OFDM systems having $N = 16$ subcarriers.\par
Fig. \ref{f9} evaluates the effects of the CIR length on the achievable rate upon adjusting the number of non-zero taps of the RIS-UE links. It can be observed that as the channel gain disperses in the delay domain, it is impossible to align all power components to the desired taps. As a result, the achievable rate decreases with the CIR length, bearing in mind that we are considering the normalized channel gain. Additionally, when considering a single non-zero tap of the RIS-UE link, the end-to-end OFDM channel relying on the conventional RIS remains a frequency-selective channel due to the diverse delays caused by distributed RISs. By contrast, with the employment of DAM, the proposed scheme converts the end-to-end channel into a frequency-flat channel by aligning all taps in the delay domain. Thus the equal power allocation solution shares the optimal performance with the WF solution. With the increase in the number of channel taps, the WF solution has a slight performance improvement compared to the equal power allocation solution. In all setups considered in Fig. \ref{f9}, the optimal configuration relying on DAM maintains the optimal performance, followed by its counterpart relying on statistical RIS configuration, which still outperforms the optimal configuration without employing DAM by about $0.02\ \text{b/s/Hz}$, respectively. The optimal configuration relying on DAM even achieves the twice achievable rate of that without DAM. Finally, the random phase configuration hardly provides any significant performance boost in all considered setups.\par
Fig. \ref{f10} shows the achievable rate versus the Rician factor of the RIS-UE link. It can be seen that the achievable rate increases with the Rician factor due to the fact that more power is concentrated on the deterministic LoS path for a large Rician factor. Hence, the reflected channels via different RIS elements are more likely to be constructively superimposed on all subcarriers for improving the average achievable rate. Specifically, for a small Rician factor, the RIS configuration relying on statistical CSI bears a severe performance penalty compared to the optimal configuration relying on instantaneous CSI. With the increase of the Rician factor, the gap between the statistical RIS configuration and the optimal RIS configuration will be progressively narrowed because a large Rician factor generally results in that all reflected channels are of a single tap and thus can be coherently superimposed on all subcarriers, even relying on only the statistical CSI (i.e., \emph{\textbf{Lemma 1}}). It shows that for a benign propagation of the RIS-UE link, i.e., a strong LoS component with few local scatterers, only the statistical CSI is adequate to complete the RIS configuration incurring inappreciable performance loss. Additionally, the proposed design obtains the best performance among all considered setups. For example, when the Rician factor is ${\zeta _{k,{\rm{RU}}}} = 0\ \text{dB}$, the achievable rate of the proposed design is more than twice that without employing DAM. As further increasing the Rician factor to ${\zeta _{k,{\rm{RU}}}} = 10\ \text{dB}$, the proposed design still remains a rate advantage of $0.04\ \text{b/s/Hz}$.\par
\subsection{Impacts of Hardware Imperfections}
Next, we consider the effect of the practical DAM component's power decay versus delay on the proposed scheme. We use exponential fading to characterize the power decay caused by the practical DAM component \cite{Nakanishia_2018_APL_Storage}, which is expressed as ${p_{{\rm{decay}}}} = {\eta ^{\bar \tau} }$, where $0 < \eta  \le 1$ and ${\bar \tau}$ denote the decay factor and delay, respectively. Fig. \ref{fig11}(a) compares the achievable rate of the proposed scheme considering power distortion to the traditional scheme, where we set $M=100$, ${C_0} =  - 20\ \text{dB}$, and ${f_s} = 10\ \text{MHz}$. For the sake of illustration, we only consider the optimal setup in the presence/absence of DAM. As observed from Fig. \ref{fig11}(a) that, with the increase of the decay factor, the performance of the proposed scheme is weakened due to the power attenuation incurred by the practical DAM component. Nevertheless, the proposed scheme still remains a performance improvement of about $0.2\ \text{b/s/Hz}$ for $\eta  = 0.8$, which implies that, for a moderate decay factor, the performance gain from aligning the strongest taps outweighs the penalty caused by the power attenuation introduced by RIS delay. It is worth noting that the RIS's delay for all setups in Fig. \ref{fig11}(a) is configured according to (\ref{eq27}), which implies that the achievable rate in the presence of power decay in Fig. \ref{fig11}(a) can be further improved upon jointly considering the RIS's delay and power decay. As a result, there exists a tradeoff between the DAM's delay and decay thus maximizing the achievable rate.\par
Finally, Fig. \ref{fig11}(b) evaluates the effects of channel estimation errors, where we consider the uplink pilot power of ${P_{{\rm{pilot}}}} = 10,\ 0,\ - 10\ \text{dBm}$, respectively. Moreover, the average noise power at the BS is set to $\sigma _{{\rm{BS}}}^2 = - 110\ \text{dBm}$ and the power decay factor of the practical DAM component is set to $\eta = 0.9$. The DFT-based reflection pattern is employed to minimize the channel estimates' MSE \cite{Jensen_ICASSP_2020_An}. Please refer to \cite{Yang_TC_2020_Intelligent} for more details about the channel estimation for RIS-assisted OFDM systems. As can be seen from the Fig. \ref{fig11}(b) that the channel estimation errors degrade the performance of both the proposed scheme employing DAM and the conventional scheme without DAM due to the defective passive beamforming based on imperfect CSI. Nevertheless, the proposed scheme always outperforms the traditional scheme under the same setup thanks to the delay adjustable RIS introduced. In particular, considering the practical DAM component's power decay of $\eta = 0.9$ and the imperfect CSI of ${P_{{\rm{pilot}}}} = - 10\ \text{dBm}$, the proposed scheme increases the achievable rate from $0.9\ \text{b/s/Hz}$ to $1.25\ \text{b/s/Hz}$, with a performance improvement of about $40\%$.\par
\section{Conclusion}\label{S7}
In this paper, we first presented the DAM that relies on varactor diodes, which is capable of storing and retrieving the impinging waves thus imposing an extra delay on the incident signals upon adjusting its EIT properties. Thanks to this new design degrees-of-freedom, RIS is more likely to align multiple reflected channels on all subcarriers concurrently. Following this, we constructed the rate maximization problem of the RIS-assisted OFDM systems by jointly optimizing the transmit power and the RIS reflection coefficients as well as RIS delays. Furthermore, we proposed an SDR-based AO algorithm to achieve a high-quality approximate solution for the formulated non-convex problem and conceived a low-complexity STA method upon aligning the strongest taps of all reflected channels. Finally, adequate simulations verify the advantages of the proposed algorithms over their traditional counterparts. In particular, the proposed algorithm outperforms the conventional schemes by $70\%$ in terms of the rate improvement for ideal scenarios, and even after taking into account the practical power decay and channel estimation errors, the proposed scheme retains a rate improvement of $40\%$. In summary, the proposed algorithm can significantly improve the RIS's performance adopting wideband signals and is expected to effectively address the beam squint effect if extended to MIMO systems \cite{Chen_TWC_2021_hybrid}.
\ifCLASSOPTIONcaptionsoff
  \newpage
\fi
\bibliography{An}
\end{document}